\documentclass[12pt]{article}
\usepackage[utf8]{inputenc}
\usepackage{amsmath}
\usepackage{tikz}
\usetikzlibrary{positioning}

\usepackage{array}
\usepackage{multirow}
\usepackage{booktabs} 
\usepackage{bm}
\usepackage{mathtools}
\usepackage{booktabs}
\usepackage{amssymb}
\usepackage{accents}
\usepackage{amsthm}
\usepackage[english]{babel} 
\usepackage[protrusion=true,expansion=true]{microtype} 
\usepackage{microtype} 
\usepackage{tabularx}
\usepackage[font = small,labelfont=bf,textfont=it]{caption} 
\usepackage[toc,page]{appendix}
\usepackage{dsfont}
\usepackage{booktabs} 
\usepackage{natbib}
\usepackage{setspace}
\usepackage{soul}
\usepackage{enumitem}

\usepackage{multirow}
\usepackage{algorithm}
\usepackage[noend]{algpseudocode}
\usepackage{etoolbox}
\algblock{ParFor}{EndParFor}
\algnewcommand\algorithmicparfor{\textbf{for}}
\algnewcommand\algorithmicpardo{}
\algnewcommand\algorithmicendparfor{}
\algrenewtext{ParFor}[1]{\algorithmicparfor\ #1\ }
\algrenewtext{EndParFor}{\algorithmicendparfor}
\newcolumntype{C}[1]{>{\centering\let\newline\\\arraybackslash\hspace{0pt}}m{#1}}

\usepackage{xcolor}

\newcommand{\argmin}{\operatorname*{argmin}}
\newcommand{\argmax}{\operatorname*{argmax}}

\usepackage{algorithm}
\usepackage{algpseudocode}

\usepackage{graphicx,epstopdf}
\makeatletter
\newcommand{\distas}[1]{\mathbin{\overset{#1}{\kern\z@\sim}}}%
\newsavebox{\mybox}\newsavebox{\mysim}
\newtheorem{theorem}{Theorem}[section]

\newtheorem{proposition}[theorem]{Proposition}

\newcommand{\distras}[1]{%
  \savebox{\mybox}{\hbox{\kern3pt$\scriptstyle#1$\kern3pt}}%
  \savebox{\mysim}{\hbox{$\sim$}}%
  \mathbin{\overset{#1}{\kern\z@\resizebox{\wd\mybox}{\ht\mysim}{$\sim$}}}%
}

\newcommand{\blind}{1}
\newcommand{\cbl}[1]{{\color{blue}{#1}}}

\begin{document}

\def\spacingset#1{\renewcommand{\baselinestretch}%
{#1}\small\normalsize} \spacingset{1.1}


\if1\blind
{
 \centering{\bf\Large Active Learning for a Recursive Non-Additive Emulator for Multi-Fidelity Computer Experiments}\\
  \vspace{0.3in}
  \centering{Junoh Heo and Chih-Li Sung\footnote{These authors gratefully acknowledge funding from NSF DMS 2113407 and 2338018.}\vspace{0.1in}\\
        Michigan State University\\
        }
    \date{\vspace{-7ex}}
} \fi

\if0\blind
{
  \bigskip
  \bigskip
  \bigskip
    \begin{center}
    {\Large\bf Active Learning for a Recursive Non-Additive Emulator for Multi-Fidelity Computer Experiments}
\end{center}
  \medskip
} \fi

\bigskip
\begin{abstract}
Computer simulations have become essential for analyzing complex systems, but high-fidelity simulations often come with significant computational costs. To tackle this challenge, multi-fidelity computer experiments have emerged as a promising approach that leverages both low-fidelity and high-fidelity simulations, enhancing both the accuracy and efficiency of the analysis. In this paper, we introduce a new and flexible statistical model, the \textit{Recursive Non-Additive (RNA) emulator}, that integrates the data from multi-fidelity computer experiments. Unlike conventional multi-fidelity emulation approaches that rely on an additive auto-regressive structure, the proposed RNA emulator recursively captures the relationships between multi-fidelity data using Gaussian process priors without making the additive assumption, allowing the model to accommodate more complex data patterns. Importantly, we derive the posterior predictive mean and variance of the emulator, which can be efficiently computed in a closed-form manner, leading to significant improvements in computational efficiency. Additionally, based on this emulator, we introduce four active learning strategies that optimize the balance between accuracy and simulation costs to guide the selection of the fidelity level and input locations for the next simulation run. We demonstrate the effectiveness of the proposed approach in a suite of synthetic examples and a real-world problem. An \textsf{R} package \textsf{RNAmf} for the proposed methodology is provided on CRAN.

\end{abstract}

\noindent%
\vfill

\newpage
\spacingset{1.5} 

\section{Introduction}
\label{sec:intro}
Computer simulations play a crucial role in engineering and scientific research, serving as valuable tools for predicting the performance of complex systems across diverse fields such as aerospace engineering \citep{mak2018efficient}, natural disaster prediction \citep{ma2022multifidelity}, and cell biology \citep{sung2020calibration}. However, conducting high-fidelity simulations for parameter space exploration can be demanding due to prohibitive costs. To address this challenge, \textit{multi-fidelity emulation} has emerged as a promising alternative. It leverages computationally expensive yet accurate high-fidelity simulations alongside computationally inexpensive but potentially less accurate low-fidelity simulations to create an efficient predictive model, \textit{emulating} the expensive computer code. By strategically integrating these simulations and designing multi-fidelity experiments, we can potentially improve accuracy without excessive computational resources.

The usefulness of the multi-fidelity emulation framework has driven extensive research in recent years. One popular approach is the Kennedy-O'Hagan (KO) model \citep{kennedy2000predicting}, which models a sequence of computer simulations from lowest to highest fidelity using a sequence of Gaussian process (GP) models \citep{gramacy2020surrogates,rasmussen2006gaussian}, linked by a linear auto-regressive framework. This model has made significant contributions across various fields employing multi-fidelity computer experiments (see, e.g., \citealp{patra2020multi, kuya2011multifidelity, demeyer2017surrogate}), and several recent developments, including \cite{qian2006building},  \cite{le2013bayesian},  \cite{le2014recursive},  \cite{qian2008bayesian},  \cite{perdikaris2017nonlinear},  and \cite{ji2022graphica} (among many others), have investigated modeling strategies for efficient posterior prediction and Bayesian uncertainty quantification.

\begin{figure}[t!]
\begin{center}
\includegraphics[width=1\textwidth]{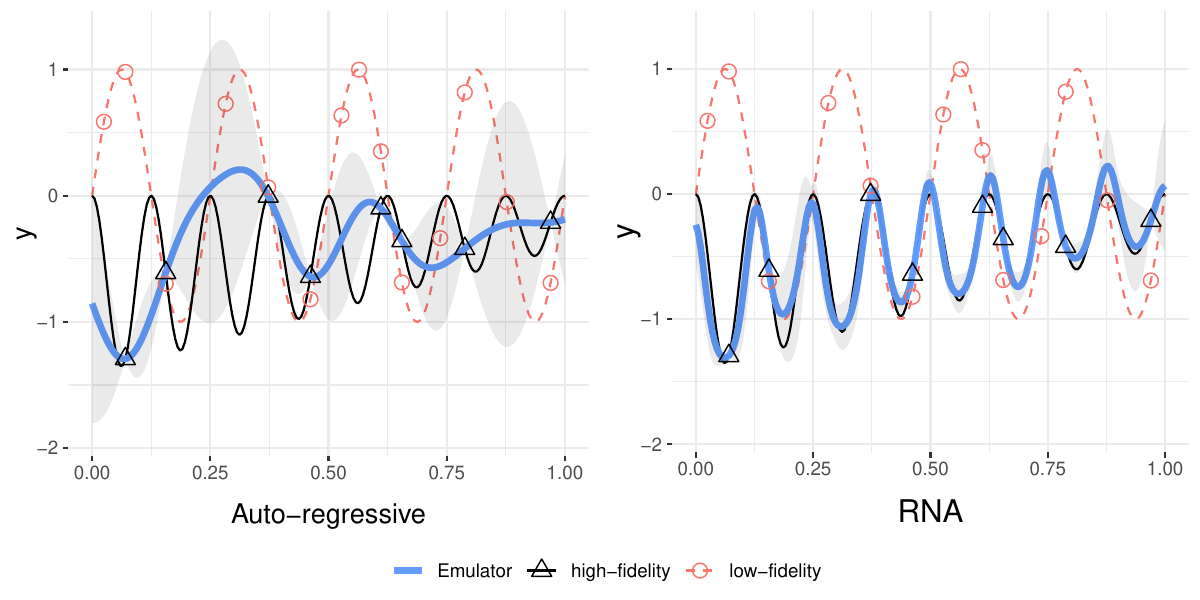} 
\end{center}
\caption{An example adapted from \cite{perdikaris2017nonlinear}, where $n_1=13$ samples (red dots) are collected from the low-fidelity simulator $f_1(x)=\sin(8\pi x)$ (red dashed line), and  $n_2=8$ samples (black triangles) are collected from the high-fidelity simulator $f_2(x)=(x-\sqrt{2})f_1(x)^2$ (black solid line). The KO emulator (left panel) and the RNA emulator (right panel) are shown as blue lines. Gray shaded regions represent the 95\% pointwise confidence intervals.}
\label{fig:demo_intro}
\end{figure}

Despite this body of work, most of these approaches rely on the assumption of linear correlation between low-fidelity and high-fidelity data, resulting in an \textit{additive} GP structure. With the growing complexity of modern data, such models face challenges in capturing complex relationships between data with different fidelity levels. As shown in the left panel of Figure \ref{fig:demo_intro}, where the relationship between high-fidelity data and low-fidelity data is nonlinear, the KO model falls short in providing accurate predictions due to its limited flexibility.

In this paper, we propose a new and flexible model that captures the \textit{nonlinear} relationships between multi-fidelity data in a \textit{recursive} manner. This flexible nonlinear functional form can encompass many existing models, including the KO model, as a special case.  Specifically, we compose GP priors to model multi-fidelity data non-additively. Hence, we refer to this proposed method as the \textit{Recursive Non-Additive (RNA) emulator}. As shown in the right panel of Figure \ref{fig:demo_intro}, the RNA emulator demonstrates superiority over the KO model by emulating the high-fidelity simulator with high accuracy and low uncertainty.

The RNA emulator belongs to the emerging field of linked/deep GP models (see, e.g., \citealp{kyzyurova2018coupling, ming2021linked, sauer2023active, ming2023deep}), where different GPs are connected in a coupled manner. To the best of our knowledge, there has been limited research on extending such results for the analysis of multi-fidelity computer experiments, and we aim to address this gap in our work. Notably, recent work by \cite{perdikaris2017nonlinear} has made progress in this direction, but their approach assumes an additive structure for the kernel function, and employs the Monte Carlo integration to handle intractable posterior distributions. Recent advancement by \cite{ko2022deep} extends deep GP models for multi-fidelity computer experiments, but still relies on the additive structure of the KO model. Similarly, \cite{cutajar2019deep} employ an additive kernel akin to \cite{perdikaris2017nonlinear} and rely on the sparse variational approximation for inference. In a similar vein, \cite{meng2020composite}, \cite{li2020multi}, \cite{meng2021multi}, and \cite{kerleguer2024bayesian} establish connections between different fidelities using (Bayesian) neural networks. 
In contrast, our proposed model not only provides great flexibility using GP priors with commonly used kernel structures  to connect  multi-fidelity data, but also provides analytical expressions for both the posterior mean and variance. This computational improvement allows for  more efficient calculations, facilitating efficient uncertainty quantification.


Leveraging this newly developed RNA emulator, we introduce four \textit{active learning} strategies to achieve enhanced accuracy while carefully managing the limited simulation resources, which is particularly crucial for computationally expensive simulations. Active learning, also known as \textit{sequential design}, involves sequentially searching for and acquiring new data points at optimal locations based on a given sampling criterion, to construct an accurate surrogate model/emulator. While active learning has been well-established for single-fidelity GP emulators \citep{gramacy2020surrogates,rasmussen2006gaussian,santner2018design}, research in the context of multi-fidelity computer experiments is scarce and more challenging. This is because it requires simultaneous selection of optimal input locations and fidelity levels, accounting for their respective simulation costs. 
Although some recent works have considered cost in specific cases like single-objective unconstrained optimization \citep{huang2006sequential,swersky2013multi,he2017optimization} and global approximation \citep{xiong2013sequential,le2015cokriging,stroh2022sequential,ehara2021,sung2022stacking}, most of these works were developed based on the KO model. Active learning for non-additive GP models has not been fully explored in the literature. In addition, popular sampling criteria for global approximation, such as ``Active Learning MacKay'' (\citealp{mckay2000comparison}, ALM) and ``Active Learning Cohn'' (\citealp{cohn1993neural}, ALC), 
remain largely unexplored in the context of multi-fidelity computer experiments. Recent successful applications of these sampling criteria to other learning problems can be found in  \cite{binois2019replication}, \cite{park2023active}, \cite{koermer2023active}, and \cite{sauer2023active}.

Our main contribution lies in advancing active learning with these popular sampling criteria, based on this newly developed RNA emulator. 
It is important to note that few existing works in the multi-fidelity deep GP literature \citep{perdikaris2017nonlinear,cutajar2019deep,ko2022deep} delve into active learning, mainly due to computational complexities associated with computing acquisition functions. In contrast, our closed-form posterior mean and variance of the RNA emulator not only facilitate efficient computation of these sampling criteria, but also provide valuable mathematical insights into the active learning.  
To facilitate broader usage, we  implement an \textsf{R} \citep{R2018} package called \textsf{RNAmf}, 
which is available on CRAN.

The structure of this article is as follows. In Section \ref{sec:review}, we provide a brief review of the KO model. Our proposed RNA emulator is introduced in Section \ref{sec:rnaemulator}. Section \ref{sec:activelearning} outlines our active learning strategies based on the RNA emulator. Numerical and real data studies are presented in Sections 5 and 6, respectively. Lastly, we conclude the paper in Section \ref{sec:conclusion}.

\section{Background}\label{sec:review}

\subsection{Problem Setup}

Let $f_l(\mathbf{x})$ represent the scalar simulation output of the computer code with input parameters $\mathbf{x}\in\Omega\subseteq\mathbb{R}^d$ at fidelity level $l=1,\ldots,L$. We assume that $L$ distinct fidelity levels of simulations are conducted for training an emulator, where a higher fidelity level corresponds to a simulator with more accurate outputs but also higher simulation costs per run.

Our primary objective is to construct an efficient emulator for the highest-fidelity simulation code, $f_L(\mathbf{x})$. For each fidelity level $l$, we perform simulations at $n_l$ design points denoted by $\mathcal{X}_l = \{\mathbf{x}_i^{[l]}\}_{i=1}^{n_l}$. These simulations yield the corresponding simulation outputs $\mathbf{y}_l := (f_l(\mathbf{x}))_{\mathbf{x} \in \mathcal{X}_l}$, representing the vector of outputs for $f_l(\mathbf{x})$ at design points $\mathbf{x} \in\mathcal{X}_l$, and each element of $\mathbf{y}_l$ is denoted by $y^{[l]}_i=f_l(\mathbf{x}^{[l]}_i)$. We assume that the  designs $\mathcal{X}_l$ are sequentially nested, i.e.,
\begin{equation}
\mathcal{X}_L\subseteq \mathcal{X}_{L-1}\subseteq\cdots\subseteq \mathcal{X}_1\subseteq \Omega,
\label{eq:nested}
\end{equation}
and $\mathbf{x}^{[l]}_i=\mathbf{x}^{[l-1]}_i$ for $i=1,\ldots,n_l$.  In other words, design points run for a higher-fidelity simulator are contained within the design points run for a lower-fidelity simulator. This property has been shown to lead to more efficient inference in various multi-fidelity emulation approaches \citep{qian2009nested, qian2009construction, haaland2010approach}. 

Furthermore, we let $C_l$ denote the simulation cost (e.g., in CPU hours) for a single run of the simulator at fidelity level $l$. Since higher-fidelity simulators are more computationally demanding, this implies that $0<C_1< C_2<\ldots< C_L$.

\subsection{Auto-regressive model}\label{sec:autoregressive}
One of the prominent approaches for modeling $f_L(\mathbf{x})$ is the KO model (also known as co-kriging model) proposed by \cite{kennedy2000predicting}, which  can be expressed in an auto-regressive form as follows:
\begin{align}\label{eq:autoregressive}
\begin{cases}
 &f_1(\mathbf{x}) = Z_1(\mathbf{x}), \\
 &f_l(\mathbf{x}) = \rho_{l-1} f_{l-1}(\mathbf{x}) + Z_l(\mathbf{x}),\quad \text{for}\quad 2\leq l\leq L,
\end{cases}  
\end{align}  
where $\rho_{l-1}$ is an unknown auto-regressive parameter, and $Z_l=(f_l- \rho_{l-1} f_{l-1})$ represents the \textit{discrepancy} between the $(l-1)$-th and $l$-th code. The KO model considers a probabilistic surrogate model by assuming that $\{Z_l\}^{L}_{l=1}$ follow independent zero-mean GP models:
\begin{equation}\label{eq:AR_GP}
Z_l(\mathbf{x}) \overset{indep.}{\sim} \mathcal{GP}\{\alpha_l(\mathbf{x}),\tau^2_lK_l(\mathbf{x},\mathbf{x}')\}, \quad l = 1, \ldots, L,
\end{equation}
where $\alpha_l(\mathbf{x})$ is a mean function, $\tau^2_l$ is a variance parameter, and $K_l(\mathbf{x},\mathbf{x}')$ is a positive-definite kernel function defined on $\Omega\times\Omega$. In the original KO paper,  $\alpha_l(\mathbf{x})$ is assumed to be $h(\mathbf{x})\beta_l$, where $h(\mathbf{x})$ is a vector of $d$ regression functions. Other common choices for $\alpha_l(\mathbf{x})$ include $\alpha_l(\mathbf{x})\equiv 0$ or $\alpha_l(\mathbf{x})\equiv\mu_l$. As for the kernel function, popular choices include the squared exponential kernel and Mat\'ern kernel \citep{stein1999interpolation}. Specifically, the \textit{anisotropic} squared exponential kernel takes the form:
\begin{align}\label{eq:sqkernel}
K_l(\mathbf{x}, \mathbf{x}')=\prod^d_{j=1}\psi(x_j,x'_j;\theta_{lj})=\prod^d_{j=1}\exp \left( -\frac{ \left( x_j - x_j' \right)^2}{\theta_{lj}}  \right),
\end{align}
where $(\theta_{l1},\ldots,\theta_{ld})$ is the \textit{lengthscale} hyperparameter, indicating that the correlation decays exponentially fast in the squared distance between $\mathbf{x}$ and $\mathbf{x}'$. The GP model \eqref{eq:AR_GP}, combined with the auto-regressive model \eqref{eq:autoregressive}, implies that, conditional on the parameters $\tau_l^2,\theta_{lj}$, and $\mu(\cdot)$, the joint distribution of $(\mathbf{y}_1,\ldots,\mathbf{y}_L)$ follows a multivariate normal distribution, and these unknown parameters can be estimated via maximum likelihood estimation or Bayesian inference.  Given the data $(\mathbf{y}_1,\ldots,\mathbf{y}_L)$, it can be shown that the posterior distribution of $f_L(\mathbf{x})$ is  also a GP. The posterior mean function can then be used to emulate the expensive simulator, while the posterior variance function can be employed to quantify the emulation uncertainty. Refer to \cite{kennedy2000predicting} for further details.

The auto-regressive framework of the KO model has led to the development of several variants.  For instance, \cite{le2014recursive} introduce a faster algorithm based on a \textit{recursive} formulation for computing the posterior of $f_l(\mathbf{x})$ more efficiently. To enhance the model's flexibility, \cite{qian2006building}, \cite{le2014recursive}, and \cite{qian2008bayesian} allow the auto-regressive parameter $\rho_l$ to depend on the input $\mathbf{x}$, that is, $f_l(\mathbf{x}) = \rho_{l-1}(\mathbf{x}) f_{l-1}(\mathbf{x}) + Z_l(\mathbf{x})$ for $2\leq l\leq L$, where the first two assume $\rho_{l-1}(\mathbf{x})$ to be a linear function, while the last one assumes $\rho_{l-1}(\mathbf{x})$ to be a GP.

\section{Recursive Non-Additive (RNA) emulator}\label{sec:rnaemulator}
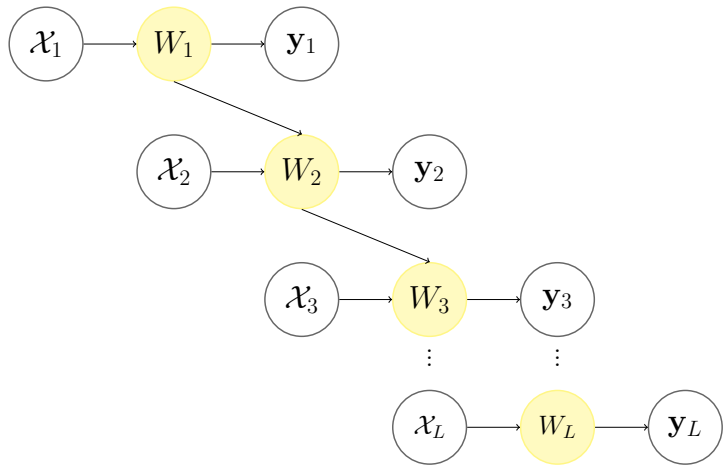
\begin{figure}[ht!]
\centering
\scalebox{0.7}{%
\begin{tikzpicture}[ 
roundnode/.style={circle, draw=black!60, fill=white!5, thick, minimum size=14mm},
roundnode2/.style={circle, draw=yellow!60, fill=yellow!30, thick, minimum size=14mm},
roundnode3/.style={circle, draw=green!60, fill=green!30, very thick, minimum size=14mm},
squarednode/.style={rectangle, draw=black!60, fill=white!5, thick, minimum size=14mm},
]
\node[roundnode,font = {\Large}]  (x_1)   {$\mathcal{X}_1$};
\node[roundnode2,font = {\Large}]    (W_1)   [right=of x_1] {$W_1$};
\node[roundnode,font = {\Large}]    (f_1)   [right=of W_1] {$\mathbf{y}_1$};
\node[roundnode,font = {\Large}]  (x_2)   [below=of W_1] {$\mathcal{X}_2$};
\node[roundnode2,font = {\Large}]    (W_2)   [right=of x_2] {$W_2$};
\node[roundnode,font = {\Large}]    (f_2)   [right=of W_2] {$\mathbf{y}_2$};

\node[roundnode,font = {\Large}]  (x_3)   [below=of W_2] {$\mathcal{X}_3$};
\node[roundnode2,font = {\Large}]    (W_3)   [right=of x_3] {$W_3$};
\node[roundnode,font = {\Large}]    (f_3)   [right=of W_3] {$\mathbf{y}_3$};

\node[below of=f_3,font = {\Large}] (dots) {\vdots};
\node[below of=W_3,font = {\Large}] (dots2) {\vdots};
\node[roundnode,font = {\large}]    (x_s)   [below=of W_3] {$\mathcal{X}_L$};
\node[roundnode2,font = {\large}] (W_s)   [right=of x_s] {$W_L$};
\node[roundnode,font = {\Large}]  (f_s)   [right=of W_s] {$\mathbf{y}_L$};
\draw[->] (x_1.east) -- (W_1.west);
\draw[->] (W_1.east) -- (f_1.west);
\draw[->] (W_1.south) -- (W_2.north);
\draw[->] (x_2.east) -- (W_2.west);
\draw[->] (W_2.east) -- (f_2.west);
\draw[->] (W_2.south) -- (W_3.north);
\draw[->] (x_3.east) -- (W_3.west);
\draw[->] (W_3.east) -- (f_3.west);
\draw[->] (x_s.east) -- (W_s.west);
\draw[->] (W_s.east) -- (f_s.west);

\end{tikzpicture}%
  }
\caption{An illustration of the recursive structure of the RNA model.}
\label{fig:modelstructure}
\end{figure}

Despite the advantages of the KO model, it results in an \textit{additive} GP model based on \eqref{eq:autoregressive} and \eqref{eq:AR_GP}, which may not adequately capture the \textit{nonlinear} relationships between data at different fidelity levels.  To overcome this limitation and achieve a more flexible representation, we propose a novel Recursive Non-Additive (RNA) emulator:
\begin{align}
\begin{cases}\label{eq:RNA}
 &f_1(\mathbf{x}) = W_1(\mathbf{x}), \\
 &f_l(\mathbf{x}) = W_l(\mathbf{x}, f_{l-1}(\mathbf{x})), \quad\text{for}\quad l=2,\ldots,L.
\end{cases}  
\end{align}
The model structure is illustrated in Figure \ref{fig:modelstructure}. This RNA model offers greater flexibility and can encompass many existing models as special cases. 
For instance, the auto-regressive KO model can be represented in the form of \eqref{eq:RNA} by setting $W_l(\mathbf{x}, f_{l-1}(\mathbf{x})) = \rho_{l-1} f_{l-1}(\mathbf{x}) + \delta(\mathbf{x})$. Similarly, the model in \cite{qian2006building}, \cite{le2014recursive}, and \cite{qian2008bayesian} can be considered a special case by setting $W_l(x, f_{l-1}(\mathbf{x})) = \rho_{l-1}(\mathbf{x}) f_{l-1}(\mathbf{x}) + \delta(\mathbf{x})$.

We model the relationship $W_l$ using a GP prior: $W_1(\mathbf{x})\sim\mathcal{GP}\{\alpha_1(\mathbf{x}),\tau^2_1K_1(\mathbf{x},\mathbf{x}')\}$ and 
\begin{equation}\label{eq:RNA_GP}
W_l(\mathbf{z}) \sim \mathcal{GP}\{\alpha_l(\mathbf{z}),\tau^2_lK_l(\mathbf{z},\mathbf{z}')\}, \quad l = 2, \cdots, L,
\end{equation}
where $\mathbf{z}=(\mathbf{x},y)$, forming a vector of size $(d+1)$, and $K_l$ is a positive-definite kernel. We consider a constant mean, i.e., $\alpha_1(\mathbf{x})=\alpha_1$ and $\alpha_l(\mathbf{z})=\alpha_l$ for $l\geq 2$. We adopt popular kernel choices for $K_l$, such as the squared exponential kernel and Mat\'ern kernel. In particular, the squared exponential kernel $K_l$ for $l\geq 2$, following \eqref{eq:sqkernel}, can be expressed as:
\begin{align}\label{eq:sqkernel_RNA}
K_l(\mathbf{z}, \mathbf{z}')&=\psi(y_{l-1},y_{l-1}';\theta_{ly})\prod^d_{j=1}\psi(x_j,x'_j;\theta_{lj})\nonumber \\
&=\exp \left( -\frac{ \left( y_{l-1} - y_{l-1}' \right)^2}{\theta_{ly}}  \right)\prod^d_{j=1}\exp \left( -\frac{ \left( x_j - x_j' \right)^2}{\theta_{lj}}  \right),
\end{align} 
where the lengthscale hyperparameter, $\bm{\theta}_l=(\theta_{l1},\ldots,\theta_{ld},\theta_{ly})$, represents a vector of size $(d+1)$,  and $K_1(\mathbf{x},\mathbf{x}')$ takes the form of \eqref{eq:sqkernel}. The Mat\'ern kernel can be similarly constructed, which is given in the Supplementary Materials \ref{app:matern}.

Combining the GP model \eqref{eq:RNA_GP} with the recursive formulation \eqref{eq:RNA} and assuming the nested design as in \eqref{eq:nested}, the observed simulations $\mathbf{y}_l$ follow a multivariate normal distribution:
$$
\mathbf{y}_l \sim \mathcal{N}_{n_l}(\alpha_l\mathbf{1}_{n_l},\tau^2_l\mathbf{K}_l)\quad\text{for}\quad l=1,\ldots,L,
$$
where $\mathbf{1}_{n_l}$ is a unit vector of size $n_l$,  $\mathbf{K}_l$ is an $n_l\times n_l$ matrix with each element $(\mathbf{K}_1)_{ij}=K_1(\mathbf{x}^{[1]}_i,\mathbf{x}^{[1]}_j)$, and $(\mathbf{K}_l)_{ij}=K_l(\mathbf{z}^{[l]}_i,\mathbf{z}^{[l]}_j)$ for $l\geq 2$, where $\mathbf{z}^{[l]}_i=(\mathbf{x}^{[l]}_i,f_{l-1}(\mathbf{x}^{[l]}_i))$. Note that $f_{l-1}(\mathbf{x}^{[l]}_i)=y^{[l-1]}_i$ due to the nested design assumption, which is the $i$-th simulation output at level $l-1$.  The parameters $\{\alpha_l,\tau^2_l,\bm{\theta}_l\}^L_{l=1}$ can be estimated by maximum likelihood estimation. Specifically, the log-likelihood (up to an additive constant) is
$$
-\frac{1}{2}\left(n_l\log(\tau_l^2)+\log(\det(\mathbf{K}_l))+\frac{1}{\tau^2_l}(\mathbf{y}_l-\alpha_l\mathbf{1}_{n_l})^T\mathbf{K}_l^{-1}(\mathbf{y}_l-\alpha_l\mathbf{1}_{n_l})\right).
$$
The parameters can then be efficiently estimated by maximizing the log-likelihood via an optimization algorithm, such as quasi-Newton optimization method of \cite{byrd1995limited}.

It is important to acknowledge that the idea of a recursive GP was previously proposed by \cite{perdikaris2017nonlinear}, referred to as a \textit{nonlinear auto-regressive model} therein. However, there are two key distinctions between their model and ours. The first distinction lies in the kernel assumption, where they assume an additive form of the kernel to better capture the auto-regressive nature. Specifically, they use   $K_l(\mathbf{z},\mathbf{z}')=K_{l1}(\mathbf{x}, \mathbf{x}')K_{l2}(f_{l-1}(\mathbf{x}), f_{l-1}(\mathbf{x}'))+K_{l3}(\mathbf{x}, \mathbf{x}')$ with valid kernel functions $K_{l1}$, $K_{l2}$, and $K_{l3}$. While our kernel function shares some similarities, particularly the first component $K_{l1}(\mathbf{x}, \mathbf{x}')K_{l2}(f_{l-1}(\mathbf{x}), f_{l-1}(\mathbf{x}'))$, the role of the second component $K_{l3}$ in predictions remains unclear. The inclusion of this component introduces $d$ hyperparameters for an anisotropic kernel, making the estimation more challenging, especially for high-dimensional problems. In contrast, we adopt the natural form of popular kernel choices, such as the squared exponential kernel in \eqref{eq:sqkernel_RNA} and the Mat\'ern kernel, placing our model within the emerging field of linked/deep GP models, which has shown promising results in the computer experiment literature \citep{kyzyurova2018coupling,ming2021linked,sauer2023active}. The second distinction is in the computation for the posterior of $f_L(\mathbf{x})$. Specifically, their model relies on Monte Carlo (MC) integration for their computation, which can be quite computationally demanding, especially in this recursive formulation. In contrast, with these popular kernel choices, we can derive the posterior mean and variance of $f_L(\mathbf{x})$ in a closed form, which is presented in the following proposition, enabling more efficient predictions and uncertainty quantification. 

The derivation of the posterior follows these steps. Starting with the GP assumption \eqref{eq:RNA_GP}, and utilizing the properties of conditional multivariate normal distribution, the posterior distribution of $f_l$ given $\mathbf{y}_l$ and $f_{l-1}$ at a new input location $\mathbf{x}$ is normally distributed, namely:
$f_1(\mathbf{x})|\mathbf{y}_1\sim\mathcal{N}(\mu_1(\mathbf{x}),\sigma_1^2(\mathbf{x}))$ 
with 
\begin{align}
\mu_1(\mathbf{x})&=\alpha_1\mathbf{1}_{n_1}+\mathbf{k}_1(\mathbf{x})^T\mathbf{K}^{-1}_1(\mathbf{y}_1-\alpha_l\mathbf{1}_{n_1}),\quad\text{and}\label{eq:gppostmean}\\
&\sigma^2_1(\mathbf{x})=\tau^2_1(1-\mathbf{k}_1(\mathbf{x})^T\mathbf{K}^{-1}_1\mathbf{k}_1(\mathbf{x})),\label{eq:gppostvar}
\end{align}
and 
$
f_l(\mathbf{x})|\mathbf{y}_l,f_{l-1}(\mathbf{x})\sim\mathcal{N}(\mu_l(\mathbf{x},f_{l-1}(\mathbf{x})),\sigma_l^2(\mathbf{x},f_{l-1}(\mathbf{x})))
$
for $l=2,\ldots,L$ with 
\begin{align}
\mu_l(\mathbf{x},f_{l-1}(\mathbf{x}))&=\alpha_l\mathbf{1}_{n_l}+\mathbf{k}_l(\mathbf{x},f_{l-1}(\mathbf{x}))^T\mathbf{K}^{-1}_l(\mathbf{y}_l-\alpha_l\mathbf{1}_{n_l}),\quad\text{and}\label{eq:gppostmean2}\\
\sigma^2_l(\mathbf{x},f_{l-1}(\mathbf{x}))&=\tau^2_l(1-\mathbf{k}_l(\mathbf{x},f_{l-1}(\mathbf{x}))^T\mathbf{K}^{-1}_l\mathbf{k}_l(\mathbf{x},f_{l-1}(\mathbf{x}))),\label{eq:gppostvar2}
\end{align}
where $\mathbf{k}_1(\mathbf{x})$ and  $\mathbf{k}_l(\mathbf{x},f_{l-1}(\mathbf{x}))$ are an $n_l\times 1$ matrix with each element $(\mathbf{k}_1(\mathbf{x}))_{i,1}=K_1(\mathbf{x}, \mathbf{x}^{[1]}_i)$ and $(\mathbf{k}_l(\mathbf{x},f_{l-1}(\mathbf{x})))_{i,1}=K_l((\mathbf{x},f_{l-1}(\mathbf{x})), (\mathbf{x}_i^{[l]},y^{[l-1]}_i))$ for $l\geq 2$, respectively. The posterior distribution of $f_l$ can then be obtained by
\begin{align*}
&p(f_l(\mathbf{x})|\mathbf{y}_1,\ldots,\mathbf{y}_l)\\=&\int\cdots\int p(f_l(\mathbf{x})|\mathbf{y}_l,f_{l-1}(\mathbf{x}))p(f_{l-1}(\mathbf{x})|\mathbf{y}_{l-1},f_{l-2}(\mathbf{x}))\cdots p(f_1(\mathbf{x})|\mathbf{y}_1) \text{d}(f_{l-1}(\mathbf{x}))\ldots \text{d}(f_1(\mathbf{x})).
\end{align*}
This posterior is analytically intractable but can be numerically approximated using MC integration, as done in \cite{perdikaris2017nonlinear}, which involves sequential sampling from the normal distribution $p(f_l(\mathbf{x})|\mathbf{y}_l,f_{l-1}(\mathbf{x}))$ from $l=1$ to $l=L$. However, this method can be computationally demanding
, especially when the dimension of $\mathbf{x}$ and the number of fidelity levels increase. To address this, we derive recursive closed-form expressions for the posterior mean and variance under popular kernel choices as follows.

\begin{proposition}\label{prop:closedform}
Under the squared exponential kernel function \eqref{eq:sqkernel_RNA}, the posterior mean and variance of $f_l(\mathbf{x})$ given the data $\{\mathbf{y}_l\}^L_{l=1}$ for $l\geq 2$ can be expressed in a recursive fashion: 
\begin{align*}
\mu^*_l(\mathbf{x}):&=\mathbb{E}[f_l(\mathbf{x})|\mathbf{y}_1,\ldots,\mathbf{y}_l]\\&=\alpha_l + \sum^{n_l}_{i=1} r_i \prod_{j=1}^d \exp\left( -\frac{(x_{j}-x^{[l]}_{ij})^2}{\theta_{lj}} \right)  \frac{1}{\sqrt{1+2\frac{
\sigma^{*2}_{l-1}(\mathbf{x}) }{\theta_{ly}}}}  \exp{\left( -\frac{(y^{[l-1]}_i-\mu^*_{l-1}(\mathbf{x}))^2}{\theta_{ly}+2\sigma^{*2}_{l-1}(\mathbf{x})} \right)},
\end{align*}   
and
\begin{align}\label{eq:postvar}
\sigma^{*2}_l(\mathbf{x}):=\mathbb{V}&[f_l(\mathbf{x})|\mathbf{y}_1,\ldots,\mathbf{y}_l]
=\tau^2_l - (\mu^*_l(\mathbf{x})-\alpha_l)^2 +\nonumber\\&\left( \sum_{i,k=1}^{n_l} \zeta_{ik}\left(r_i r_k - \tau^2_l (\mathbf{K}^{-1}_l)_{ik}  \right)\prod_{j=1}^d \exp{ \left(-\frac{(x_{j}-x^{[l]}_{ij})^2+(x_{j}-x^{[l]}_{kj})^2}{\theta_{lj}}\right)} \right),
\end{align}    
where $r_i = (\mathbf{K}^{-1}_l (\mathbf{y}_l-\alpha_l\mathbf{1}_{n_l}))_i$, and 
\begin{align}\label{eq:postvar_zeta}
\zeta_{ik} = \frac{1}{\sqrt{1+4\frac{\sigma^{*2}_{l-1}(\mathbf{x})}{\theta_{ly}}}}  \exp{\left( -\frac{(\frac{y^{[l-1]}_i+y^{[l-1]}_k}{2}-\mu^*_{l-1}(\mathbf{x}))^2}{\frac{\theta_{ly}}{2}+2\sigma^{*2}_{l-1}(\mathbf{x})} -\frac{(y^{[l-1]}_i-y^{[l-1]}_k)^2}{2\theta_{ly}} \right)}.
\end{align}
For $l=1$, it follows that $\mu^*_1(\mathbf{x})=\mu_1(\mathbf{x})$ and $\sigma^{*2}_1(\mathbf{x})=\sigma^{2}_1(\mathbf{x})$ as in \eqref{eq:gppostmean} and \eqref{eq:gppostvar}, respectively.
\end{proposition}

The posterior mean and variance under a Mat\'ern kernel with the smoothness parameter $\nu=1.5$ and $\nu=2.5$ are provided in the Supplementary Materials \ref{supp:maternposterior}, and the detailed derivations for Proposition \ref{prop:closedform} are provided in Supplementary Materials \ref{supp:proof3.1}, which follow the proof of \cite{kyzyurova2018coupling} and \cite{ming2021linked}. With this proposition, the posterior mean and variance can be efficiently computed in a recursive fashion. Similar to \cite{kyzyurova2018coupling} and \cite{ming2021linked}, we adopt the \textit{moment matching} method, using a Gaussian distribution to approximate the posterior distribution with the mean and variance presented in the proposition. The parameters in the posterior distribution, including $\{\alpha_l,\tau^2_l,\bm{\theta}_l\}^L_{l=1}$, can be plugged in by their estimates.

Notably, Proposition \ref{prop:closedform} can be viewed as a simplified representation of Theorem 3.3 from \cite{ming2021linked} for constructing a linked GP surrogate. However, it is important to highlight the distinctions and contributions of our work, particularly in the context of multi-fidelity computer experiments. Firstly, there are currently no existing closed-form expressions for the posterior mean and variance in the multi-fidelity deep GP literature. By providing such expressions, our work fills this gap, offering valuable mathematical insights and enhancing computational efficiency for active learning strategies, which will be discussed in Section \ref{sec:insightemulator} and Section \ref{sec:activelearning}. Additionally, while the linked GP model provides a general framework, much of the discussion in their work focuses on \textit{sequential} GPs, where the output of the high-layer emulator depends solely on the output of the low-layer emulator, i.e., $W_2(W_1(\mathbf{x}))$. Our setup differs slightly, as the high-fidelity emulator in our RNA framework depends not only on the output of the low-fidelity emulator but also on the input variables directly, i.e., $W_2(\mathbf{x},W_1(\mathbf{x}))$. This difference in formulation is important and impacts the design of active learning strategies in our framework.

Similar to conventional GP emulators for single-fidelity deterministic computer models, the proposed RNA emulator also exhibits the interpolation property, which is described in the following proposition. The proof is provided in Supplementary Materials \ref{app:proof3.2}.

\begin{proposition}\label{prop:interpolation}
The RNA emulator satisfies interpolation property, that is, $ \mu^*_l(\mathbf{x}_i^{[l]}) = y_i^{[l]}$, and $\sigma^{*2}_l(\mathbf{x}_i^{[l]}) = 0$, where $\{(\mathbf{x}_i^{[l]},y_i^{[l]})\}_{i=1,\ldots,n_l}$ are the training samples.
\end{proposition}

An example of this posterior distribution is presented in the right panel of Figure \ref{fig:demo_intro},  illustrating that the posterior mean closely aligns with the true function, and the confidence intervals constructed by the posterior variance cover the true function. For further insights into how this nonlinear relationship modeling can effectively reconstruct the high-fidelity function $f_2(x)$ for this example, we refer to \cite{perdikaris2017nonlinear}.

Our \textsf{R} package,  \textsf{RNAmf}, implements the parameter estimation and computations for the closed-form posterior mean and variance using a squared exponential kernel and a Mat\'ern kernel with smoothness parameters of $1.5$ and $2.5$. 

\subsection{Insights into the RNA emulator}\label{sec:insightemulator}

We delve into the RNA emulator, exploring its mathematical insights and investigating scenarios where this method may succeed or encounter challenges. 

For the sake of simplicity in explanation, we consider two fidelity levels ($L=2$) and assume the input $x$ is one-dimensional. According to Proposition \ref{prop:closedform}, under a squared exponential kernel function,  the RNA emulator yields the following posterior mean:
\begin{align*}
\mu^*_2(x)=\alpha_2+\sqrt{\frac{\theta_{2y}}{\theta_{2y}+2
\sigma^{*2}_{1}(x) }}\sum^{n_2}_{i=1} r_i  \exp\left( -\frac{(x-x^{[2]}_{i})^2}{\theta_{2}}  -\frac{(y^{[1]}_i-\mu^*_{1}(x))^2}{\theta_{2y}+2\sigma^{*2}_{1}(x)} \right),
\end{align*}  
where $r_i = (\mathbf{K}^{-1}_2 (\mathbf{y}_2-\alpha_2\mathbf{1}_{n_2}))_i$.

The mathematical expression reveals several insights into the behavior of the RNA emulator. Firstly, it reveals the impact of the uncertainty in the low-fidelity model, $\sigma^{*2}_{1}(x)$, on the posterior mean $\mu^*_2(x)$. In scenarios where $\sigma^{*2}_{1}(x)=0$ for all $x\in\Omega$, $\mu^*_2(x)$ mirrors the posterior mean when $\mu^*_1(x)$ is replaced with the true low-fidelity function $f_1(x)$. Consequently, the term $\sqrt{\frac{\theta_{2y}}{\theta_{2y}+2 \sigma^{*2}_{1}(x) }}$ acts as a scaling factor for the posterior mean, adjusting the influence of the uncertainty $\sigma^{*2}_{1}(x)$ on the overall prediction to account for the approximation error between $\mu^*_1(x)$ and $f_1(x)$. Additionally, the inflated denominator of $\frac{(y^{[1]}_i-\mu^*_{1}(x))^2}{\theta_{2y}+2\sigma^{2}_{1}(x)}$ by the low-fidelity model uncertainty also aids in mitigating the approximation error, indicating a slower decay in correlation with the squared distance between the low-fidelity observations $y^{[1]}_i$ and $\mu^*_{1}(x)$. Both aspects ensure a balanced integration of high and low-fidelity information, which is particularly crucial when dealing with limited samples from low-fidelity data. 

\begin{figure}[t!]
\begin{center}
\includegraphics[width=0.9\textwidth]{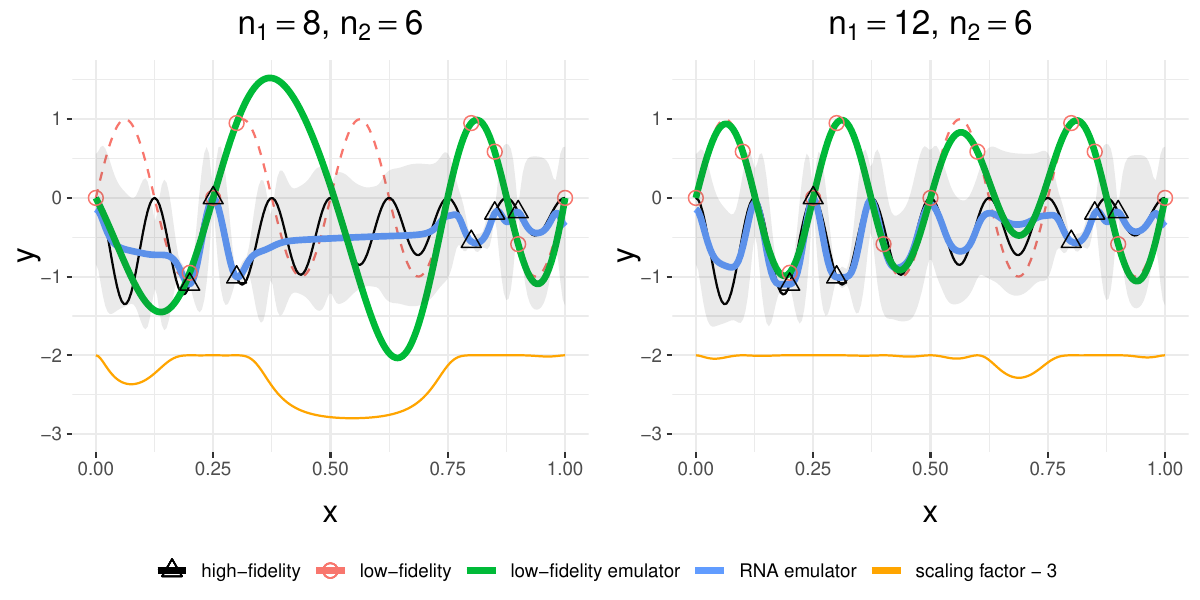} 
\end{center}
\caption{Illustration of RNA emulator insights using the Perdikaris example. The left panel and right panel depict results obtained with different sample sizes of low-fidelity data (red dots), $n_1=8$ (left) and $n_1=12$ (right), alongside the same high-fidelity data (black triangles) of size $n_2=6$. The scaling factor is the orange solid line, with values shifted by subtracting 3.}
\label{fig:RNA_insight}
\end{figure}

Figure \ref{fig:RNA_insight} demonstrates an example of how the low-fidelity emulator impacts RNA emulation performance. The left panel illustrates that with limited low-fidelity data ($n_1=8$), especially in the absence of data at $x\in(0.3,0.8)$, the posterior mean of the low-fidelity emulator, $\mu^*_1(x)$ (represented by the green line), inaccurately predicts the true low-fidelity simulator $f_1(x)$ (red dashed line). In this scenario, the scaling factor (orange line), $\sqrt{\frac{\theta_{2y}}{\theta_{2y}+2 \sigma^{2}_{1}(x) }}$, 
is very small for those poor predictions of $\mu^*_1(x)$, particularly for $x\in(0.3,0.8)$. 
This results in $\mu^*_2(x)$ being close to the mean estimate $\hat{\alpha}_2$. This is not surprising because there is no data available from both low-fidelity and high-fidelity simulators in this region, leading to the posterior mean reverting back to the mean estimate. With an increase in low-fidelity data ($n_1=12$), which makes $\mu^*_1(x)$ much closer to the true $f_1(x)$, the scaling factor is close to one everywhere, significantly enhancing the accuracy of the RNA emulator. 

\begin{figure}[t!]
\begin{center}
\includegraphics[width=0.9\textwidth]{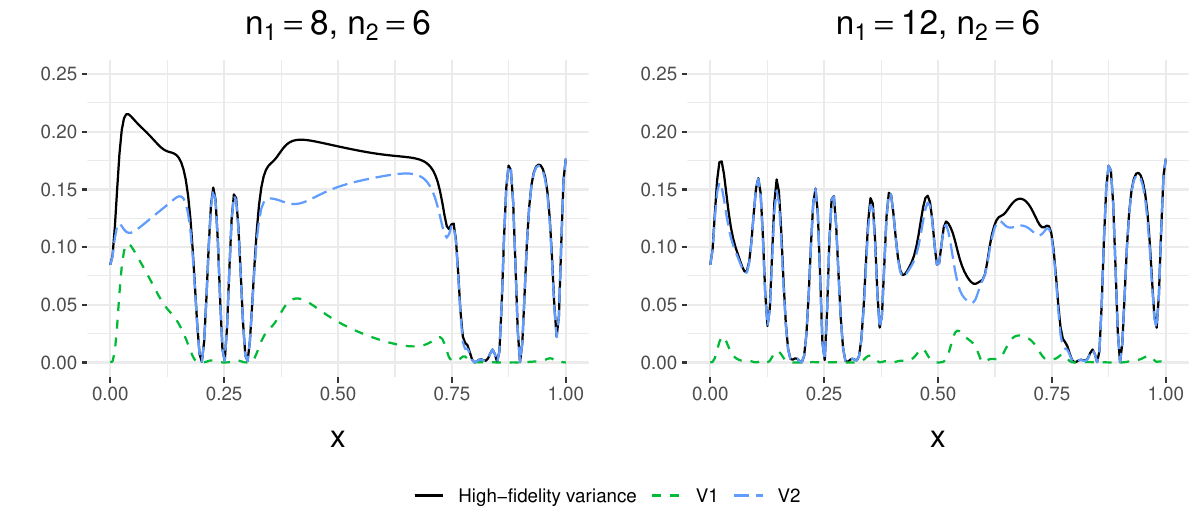} 
\end{center}
\caption{Illustration of decomposition of $\sigma^{*2}_2(x)$ (black solid line) for the examples of Figure \ref{fig:RNA_insight}, where $V_1$ is the blue dashed line and $V_2(x)$ is the green dashed line.}
\label{fig:RNAvariance_insight}
\end{figure}

The posterior variance can be written as (see Supplementary Materials S1)
\begin{equation}\label{eq:decomposition}
    \sigma^{*2}_2(x)=\mathbb{V} \left[ \mathbb{E} [f_2(x) | f_1(x), \mathbf{y}_1, \mathbf{y}_2 ] \right] + \mathbb{E} \left[\mathbb{V} [f_2(x) | f_1(x), \mathbf{y}_1, \mathbf{y}_2 ] \right],
\end{equation}
where both terms can be expressed in a closed form as in \eqref{eq:V1} and \eqref{eq:V2}, respectively.
Define $V_1(x)=\mathbb{V} \left[ \mathbb{E} [f_2(x) | f_1(x), \mathbf{y}_1, \mathbf{y}_2 ] \right]$ and $V_2(x)=\mathbb{E} \left[\mathbb{V} [f_2(x) | f_1(x), \mathbf{y}_1, \mathbf{y}_2 ] \right]$, then $V_1(x)$ represents the overall contribution of the GP emulator $W_1$ to $\sigma^{*2}_2(x)$ and $V_2(x)$ represents the contribution of the GP emulator $W_2$ to $\sigma^{*2}_2(x)$. This decomposition mirrors that of \cite{ming2021linked} within the context of linked GPs. Figure \ref{fig:RNAvariance_insight} illustrates this decomposition for the examples in Figure \ref{fig:RNA_insight}. It can be seen that for both scenarios, $V_2$ appears to dominate $V_1$, indicating that $W_2$ contributes more uncertainty than $W_1$. However, when we have limited low-fidelity data (left panel), $V_1$ exhibits a very high peak at $x\approx0.04$ with a value close to 0.10, even very close to the maximum value of $V_2$. From an active learning perspective, if the cost of evaluating $f_1(x)$ is cheaper than $f_2(x)$, then it's sensible to select the next sample from the cheaper $f_1(x)$ to reduce $\sigma^{*2}_2(x)$. On the other hand, when we have more low-fidelity data (right panel), $V_1$ remains very small everywhere compared to $V_2$, indicating that selecting the next sample from $f_2(x)$ would be more effective in reducing the predictive uncertainty. More details of active learning strategies will be introduced in the next section.

\section{Active learning for RNA emulator}\label{sec:activelearning}
We present four active learning (AL) strategies aimed at enhancing the predictive capabilities of the proposed model through the careful design of computer experiments. 
These strategies encompass the dual task of not only identifying the optimal input locations but also determining the most appropriate fidelity level. 

We suppose that an initial experiment of sample size $n_l$ for each fidelity level $l$, following a nested design $\mathcal{X}_L\subseteq\cdots\subseteq\mathcal{X}_1$, is conducted, for which a space-filling design is often considered, such as the nested Latin hypercube design \citep{qian2009nested}.  AL  seeks to optimize a selection criterion for choosing the next point $\mathbf{x}^{[l]}_{n_l+1}$ at fidelity level $l$, carrying out its corresponding simulation $y^{[l]}_{n_l+1}=f_l(\mathbf{x}^{[l]}_{n_l+1})$, and thus augmenting the dataset.

\subsection{Active Learning Decomposition (ALD)}
We first introduce an active learning criterion inspired by Section \ref{sec:insightemulator} and the variance-based adaptive design for linked GPs outlined in \cite{ming2021linked}. Specifically, we extend the decomposition of \eqref{eq:decomposition} to encompass $L$ fidelity levels:
\begin{equation}\label{eq:ALD}
    \sigma^{*2}_L(\mathbf{x})=\sum^L_{l=1}V_l(\mathbf{x}),
\end{equation}
where $V_l(\mathbf{x})$ represents the contribution of each GP emulator $W_l$ at fidelity level $l$ to $\sigma^{*2}_L(\mathbf{x})$:
$$
V_l(\mathbf{x})=\mathbb{E}\cdots\mathbb{E}\mathbb{V}\mathbb{E}\cdots \mathbb{E}\left[f_L(\mathbf{x}) | f_{L-1}(\mathbf{x}),\cdots, f_{1}(\mathbf{x}),\mathbf{y}_L,\cdots, \mathbf{y}_1 \right],
$$
with $\mathbb{V}$ being at the $l$-th term. The expectation or variance in the $l$-th term is taken with respect to the variable $f_l(\mathbf{x})$. When $L=2$, the closed-form expression for $V_l(\mathbf{x})$ is available, as shown in \eqref{eq:decomposition}. For $L=3$, each $V_l(\mathbf{x})$ can be easily approximated using MC methods. We detail the calculation of $V_l(\mathbf{x})$ for the settings of $L=2$ and $L=3$ in Supplementary Materials \ref{supp:variancedecomposition}. However, the calculation becomes more cumbersome for $L \geq 4$, which we leave as a topic for future development. 

Considering the simulation cost $C_l$, our approach guides the selection of the next point $\mathbf{x}^{[l]}_{n_l+1}$ at fidelity level $l$ by maximizing the criterion, which we refer to as Active Learning Decomposition (ALD):
$$
(l^*,\mathbf{x}_{n_{l^*}+1}^{[l^*]}) = \argmax_{l\in\{1,\ldots,L\}; \mathbf{x} \in \Omega} \frac{V_l(\mathbf{x})}{\sum^l_{j=1}C_j},
$$
which aims to maximize the ratio between each contribution $V_l(\mathbf{x})$ to $\sigma^{*2}_L(\mathbf{x})$ and the simulation cost $\sum^l_{j=1}C_j$ at each fidelity level $l$.

Simulation costs are incorporated to account for the nested structure. That is, to run the simulation $f_{l^*}(\mathbf{x}_{n_{l^*}+1}^{[l^*]})$, we also need to run $f_{l}(\mathbf{x}_{n_{l}+1}^{[l]})$ with $\mathbf{x}^{[l]}_{n_l+1}=\mathbf{x}^{[l^*]}_{n_{l^*}+1}$ for all $1\leq l<l^*$.  It is also worth mentioning that the cost can be tailored to depend on the input $\mathbf{x}$, as done in \cite{he2017optimization} and \cite{stroh2022sequential}.

\subsection{Active Learning MacKay (ALM)}\label{sec:ALM}

A straightforward but commonly used sampling criterion in AL   is to select the next point that maximizes the posterior predictive variance \citep{mackay1992information}. Extending this concept to our scenario, 
we choose  the next point by maximizing the ALM criterion:
\begin{align}\label{eq:ALM}
    (l^*,\mathbf{x}_{n_{l^*}+1}^{[l^*]}) = \argmax_{l\in\{1,\ldots,L\}; \mathbf{x} \in \Omega} \frac{\sigma^{*2}_l(\mathbf{x})}{\sum^l_{j=1}C_j}.
\end{align}
Note that after running the simulation at the optimal input location $\mathbf{x}^{[l^*]}_{n_{l^*}+1}$ at level $l^*$, the posterior predictive variance $\sigma^{*2}_{l^*}(\mathbf{x}^{[l^*]}_{n_{l^*}+1})$ becomes zero (see Proposition \ref{prop:interpolation}). In other words, our selection of the optimal level hinges on achieving the highest ratio of \textit{uncertainty reduction at $\mathbf{x}^{[l^*]}_{n_{l^*}+1}$} to the simulation cost. 

The computation of ALM criterion is facilitated by the availability of the closed-form expression of the posterior predictive variance as in \eqref{eq:postvar}, which in turn simplifies the optimization process of  \eqref{eq:ALM}. In particular, the optimal input location $\mathbf{x}^{[l]}_{n_l+1}$ for each $l$ can be efficiently obtained through the \textsf{optim} library in \textsf{R}, using the \textsf{method=L-BFGS-B} option, which performs a quasi-Newton optimization approach of \cite{byrd1995limited}.


\subsection{Active Learning Cohn (ALC)}
Another widely employed, more aggregate criterion is Active Learning Cohn (ALC) \citep{cohn1993neural,seo2000gaussian}. In contrast to ALM, ALC selects an input location that maximizes the reduction in posterior variances \textit{across the entire input space} after running this selected simulation. Extending the concept to our scenario, we choose  the next point   by maximizing the ALC criterion:
\begin{align}\label{eq:ALC}
    (l^*,\mathbf{x}_{n_{l^*}+1}^{[l^*]}) = \argmax_{l\in\{1,\ldots,L\}; \mathbf{x} \in \Omega} \frac{\Delta \sigma_L^{2}(l,\mathbf{x})}{\sum^l_{j=1}C_j},
\end{align}
where $\Delta \sigma_L^{2}(l,\mathbf{x})$ is the \textit{average  reduction in variance} (of the highest-fidelity emulator) from the current design measured through a choice of the fidelity level $l$ and the input location $\mathbf{x}$, augmenting the design. That is,
\begin{align}\label{eq:averagevariancereduction}
\Delta \sigma_L^{2}(l,\mathbf{x})=\int_{\Omega} \sigma_L^{*2}(\bm{\xi})-\tilde{\sigma}_L^{*2}(\bm{\xi};l,\mathbf{x}){\rm{d}}\bm{\xi},
\end{align}
where $\sigma_L^{*2}(\bm{\xi})$ is the  posterior variance  of $f_L(\bm{\xi})$ based on the current design $\{\mathcal{X}_l\}^L_{l=1}$, and $\tilde{\sigma}_L^{*2}(\bm{\xi};l,\mathbf{x})$ is the  posterior variance  based on the \textit{augmented design} combining the current design and a new input location $\mathbf{x}$ at each fidelity level lower than or equal to $l$, i.e., $\{(\mathcal{X}_1\cup\mathbf{x}_{n_1+1}^{[1]}),\ldots,(\mathcal{X}_l\cup\mathbf{x}_{n_{l}+1}^{[l]}),\mathcal{X}_{l+1},\ldots,\mathcal{X}_L\}$ with $\mathbf{x}_{n_1+1}^{[1]}=\cdots=\mathbf{x}_{n_{l}+1}^{[l]}=\mathbf{x}$. Once again, the incorporation of the new input location $\mathbf{x}$ at each fidelity level lower than $l$ is due to the nested structure assumption. In other words, our selection of the optimal level involves maximizing the ratio of \textit{average reduction in the variance of the highest-fidelity emulator} to the associated simulation cost.  In practice, the integration in \eqref{eq:averagevariancereduction} can be approximated by numerical methods, such as MC integration. 

Unlike ALM where the influence of design augmentation on the variance of the highest-fidelity emulator is unclear, ALC is specifically designed to \textit{maximize the reduction in variance of the highest-fidelity emulator}. However, the ALC strategy involves requiring knowledge of future outputs $y^{[s]}_{n_s+1}=f_s(\mathbf{x}^{[s]}_{n_s+1})$ for all $1 \leq s\leq l$, as they are involved in  $\tilde{\sigma}_L^{*2}(\bm{\xi};l,\mathbf{x})$ (as seen in \eqref{eq:postvar_zeta}), but these outputs are not available prior to conducting the simulations. A possible approach to address this issue is through MC approximation to \textit{impute} the outputs. Specifically, we can impute  $y^{[s]}_{n_s+1}$ for each $1\leq s\leq l$ by drawing samples from the posterior distribution of $f_s(\mathbf{x}^{[s]}_{n_s+1})$ based on the \textit{current} design, which is a normal distribution with the posterior mean and variance presented in Proposition \ref{prop:closedform}. This allows us to repeatedly compute $\tilde{\sigma}_L^{*2}(\bm{\xi};l,\mathbf{x})$  using the imputations and average the results to approximate the variance.  Notably, with the imputed output $y^{[s]}_{n_s+1}$, the  variance $\tilde{\sigma}_L^{*2}(\bm{\xi};l,\mathbf{x})$ can be efficiently computed using the Sherman--Morrison formula \citep{harville1998matrix} for updating the covariance matrix's inverse, $\mathbf{K}_l^{-1}$, from $\sigma_L^{*2}(\bm{\xi})$ \citep{gramacy2020surrogates}. 

In contrast to ALM, maximizing the ALC criterion \eqref{eq:ALC} can be quite computationally expensive due to the costly MC approximation to compute \eqref{eq:averagevariancereduction}. To this end, an alternative strategy is proposed to strike a compromise by combining the two criteria.

\subsection{Two-step approach: ALMC}
Given the distinct advantages and limitations of both ALM and ALC criteria (details of which are referred to Chapter 6 of \citealp{gramacy2020surrogates}), for a comprehensive exploration, we can contemplate their combination. Inspired by \cite{le2015cokriging}, we introduce a hybrid approach, which we refer to as \textit{ALMC}. First, the optimal input location is selected by maximizing the posterior predictive variance of the \textit{highest} fidelity emulator:
\begin{align*}
    \mathbf{x}^* = \argmax_{\mathbf{x} \in \Omega} \sigma^{*2}_L(\mathbf{x}).
\end{align*}
Then, the ALC criterion determines the fidelity level  with the identified input location:
\begin{align*}
    l^* = \argmax_{l\in\{1,\ldots,L\}} \frac{\Delta \sigma_L^{2}(l,\mathbf{x}^*)}{\sum^l_{j=1}C_j}.
\end{align*}
Unlike ALM, this hybrid approach focuses on the \textit{direct impact} on the highest-fidelity emulator. It first identifies the sampling location that maximizes $\sigma^{*2}_L(\mathbf{x})$, and then determines which level selection will effectively reduce the overall variance of the highest-fidelity emulator across the input space \textit{after running this location}.
This synergistic approach is not only expected to capture the advantages of both ALM and ALC, but also offers computational efficiency advantages compared to the ALC method in the previous subsection. This is due to the fact that the optimization for $\mathbf{x}^*$ by maximizing the closed-form posterior variance is computationally much cheaper, as discussed in Section \ref{sec:ALM}.

\begin{figure}[!t]
\begin{center}
\includegraphics[width=1\textwidth]{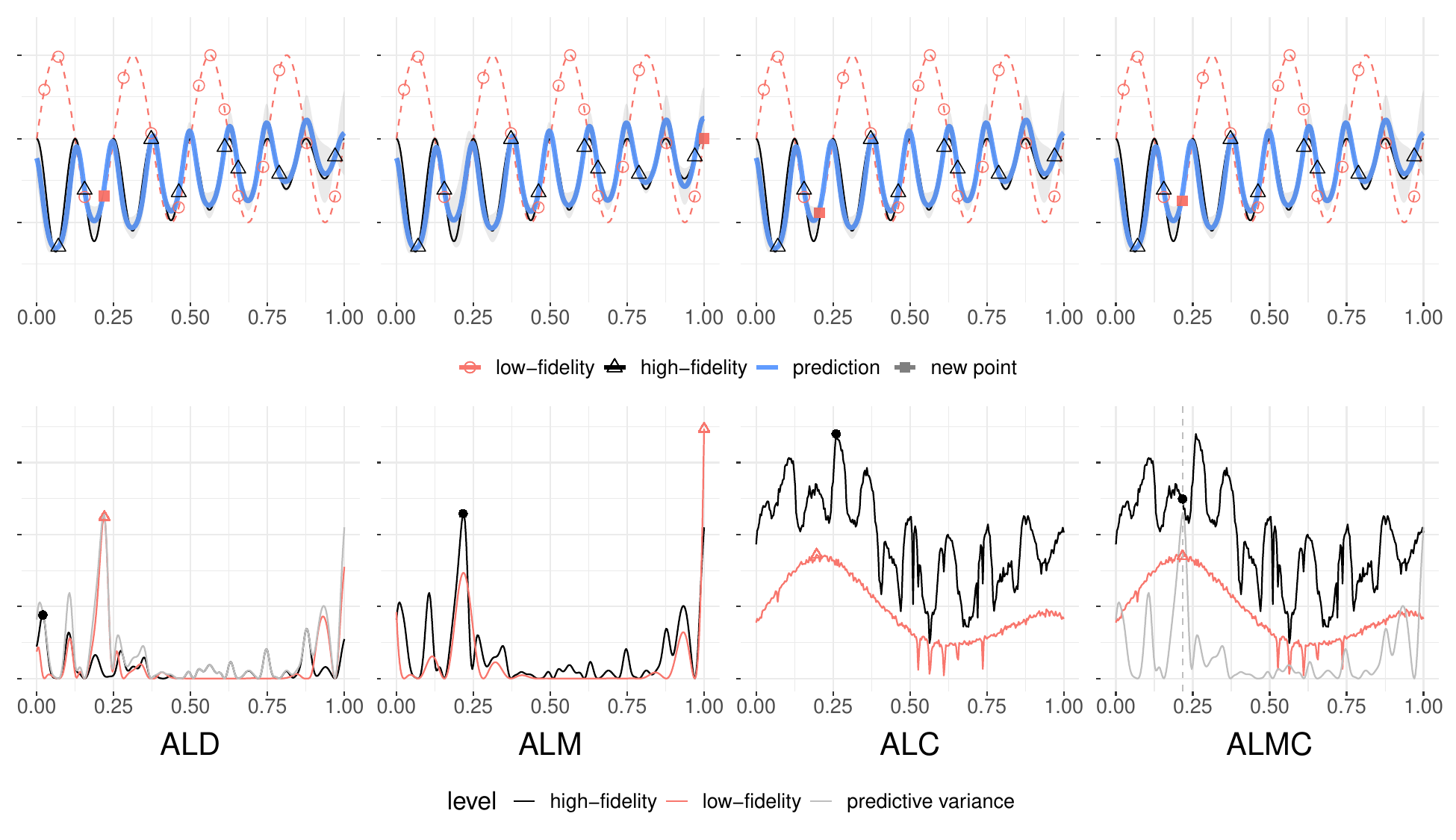}
\end{center}
\caption{Demonstration of the four active learning strategies using the example in the right panel of Figure \ref{fig:demo_intro}. The criteria of the four strategies are presented in the bottom panel, 
where the dots represent the optimal input locations for each of the simulators. Notably, ALD utilizes the gray line to illustrate $\sigma^{*2}_2(x)$, which is decomposed into $V_1(x)$ (depicted in red) and $V_2(x)$ (depicted in black). ALMC, on the other hand, employs the gray line to determine the optimal input location and then utilizes the red and black lines (which are identical to ALC) to decide the fidelity level. The upper panels show the corresponding fits after adding the selected points to the training dataset, where the solid dots represent the chosen samples, all of which select the low-fidelity simulator.}
\label{fig:AL_demo}
\end{figure}

Figure \ref{fig:AL_demo} demonstrates the effectiveness of these four strategies for the example in the right panel of Figure \ref{fig:demo_intro}. Consider the simulation costs: $C_1=1$ and $C_2=3$ for the two simulators. It shows that, for all four criteria, the choice is consistently in favor of selecting the low-fidelity simulator to augment the dataset. While the selected locations differ, ALD, ALC, and ALMC  all fall within the range of $[0.18, 0.25]$, which, as per the current design (prior to running this simulation), holds large uncertainty, as seen in the right panel of Figure \ref{fig:demo_intro}. ALM selects the sample at the boundary of the input space. All these selection outcomes contribute to an overall improvement in emulation accuracy, while simultaneously reducing global uncertainty, even when opting for low-fidelity data alone.

\subsection{Remark on the AL strategies}
In this section, we delve deeper into the merits of the AL strategies, with a focus on the conditions favoring each method. To gain deeper insights, we consider a synthetic example generated from a 2-level Currin function \citep{xiong2013sequential,kerleguer2024bayesian}, with the explicit form provided in Supplementary Materials \ref{app:functions}. Assuming simulation costs $C_1=1$ and $C_2=3$, we employ the four AL strategies until reaching a total budget of 15.

\begin{figure}[!t]
\begin{center}
\includegraphics[width=1\textwidth]{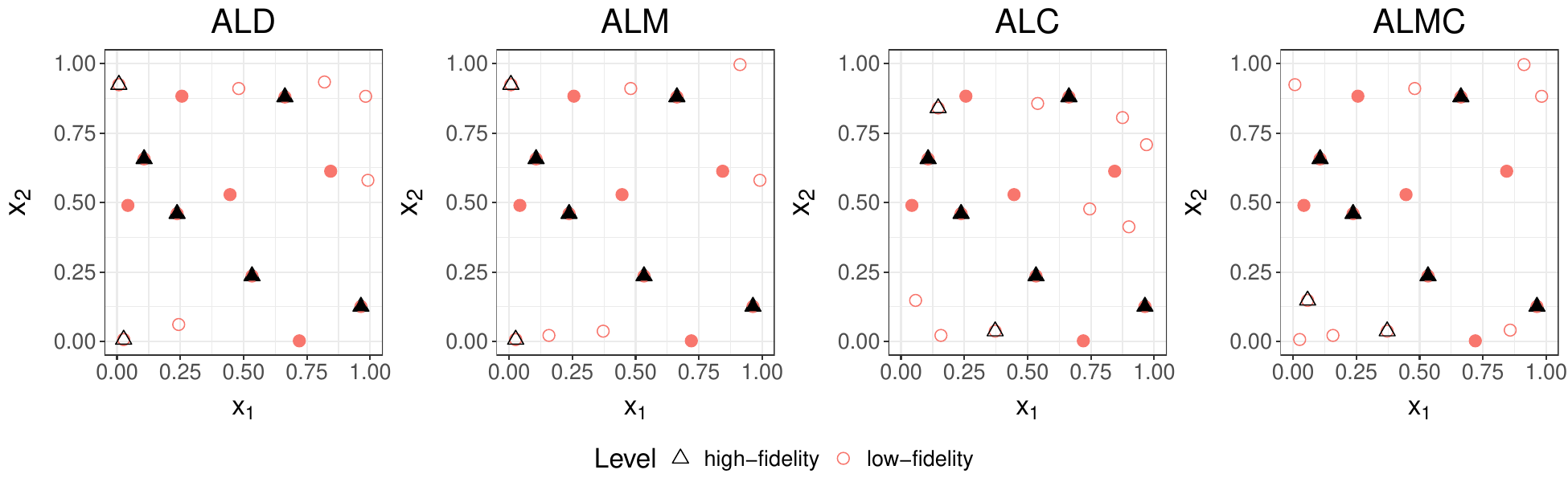} 
\end{center}
\caption{Selected input locations by four proposed strategies with a total budget of 15, where the simulation costs are $C_1=1$ and $C_2=3$. The initial design points are represented as filled shapes.}
\label{fig:AL_comparison}
\end{figure}

Figure \ref{fig:AL_comparison} showcases the selected sites within the input space $[0,1]^2$. Similar to discussions on AL for single-fidelity GPs  \citep{seo2000gaussian,gramacy2009adaptive,bilionis2012multi,beck2016sequential}, ALM tends to push selected data points towards the boundaries of the input space, whereas ALC avoids boundary locations. ALD and ALMC, inheriting attributes of ALM, exhibit similar behavior to ALM. The choice between them depends on the underlying true function: if the function in the boundary region is flat and exhibits more variability in the interior, then ALC may be preferable. Regarding computational efficiency, ALD, ALM, and ALMC benefit from closed-form expressions of the posterior variance, requiring only a few seconds per acquisition. In contrast, ALC is more computationally demanding due to extensive MC sampling efforts, taking several minutes per acquisition. 

It is worth noting that if the scale of low-fidelity outputs significantly exceeds that of high-fidelity outputs, ALM may consistently favor low-fidelity levels in the initial acquisitions, as the maximum of the low-fidelity posterior variance tends to be large. However, it's unclear whether this selection is effective, as maximizing the posterior variance of the low-fidelity emulator doesn't necessarily translate to a reduction in the uncertainty of the high-fidelity emulator. In contrast, the other three methods focus on directly impacting the high-fidelity emulator by selecting points, making them independent of the scale. In summary, considering the discussions above and the findings from our empirical studies in Sections \ref{sec:numericstudies} and \ref{sec:realdata}, ALD (for $L\leq 3$) and ALMC generally emerge as favorable choices, offering accurate RNA emulators along with computational efficiency.



\section{Numerical Studies}\label{sec:numericstudies}
In this section, we conduct a suite of numerical experiments to examine the performance of the proposed approach. The experiments encompass two main aspects. In Section \ref{sec:emulationperformance}, we assess the predictive capabilities of the proposed RNA emulator, while Section \ref{sec:ALperformance} delves into the evaluation of the performance of the proposed AL strategies. 

We consider the anisotropic squared exponential kernel as in \eqref{eq:sqkernel_RNA} for the proposed model, a choice that is also shared by our competing methods. All experiments are performed on a MacBook Pro laptop with 2.9 GHz 6-Core Intel Core i9 and 16Gb of RAM. 

\subsection{Emulation performance}\label{sec:emulationperformance}
We begin by comparing the predictive performance of the proposed RNA emulator (labeled \texttt{RNAmf}) with two other methods in the numerical experiments: the co-kriging model (labeled \texttt{CoKriging}) by \cite{le2014recursive}, and the nonlinear auto-regressive multi-fidelity GP (labeled \texttt{NARGP})  by \cite{perdikaris2017nonlinear}. The two methods are readily available through open repositories, specifically the \textsf{R} package \textsf{MuFiCokriging} \citep{RMuFiCokriging} and the \textsf{Python} package on the GitHub repository  \citep{NARGP}, respectively. Note that the multi-fidelity deep GP of \cite{cutajar2019deep}, which can be implemented using the \textsf{Python} package \texttt{emukit} \citep{emukit2019,emukit2023},  is not included in our comparison due to software limitations. We encountered challenges during implementation as the package relies on an outdated package, rendering it incompatible with our current environment.

Five synthetic examples commonly used in the literature to evaluate emulation performance in multi-fidelity simulations are considered, including the two-level Perdikaris function \citep{perdikaris2017nonlinear, kerleguer2024bayesian},
the two-level Park function \citep{park1991tuning,xiong2013sequential}, the three-level Branin function \citep{sobester2008engineering}, the two-level Borehole function \citep{morris1993bayesian,xiong2013sequential}, and the two-level Currin function \citep{xiong2013sequential, kerleguer2024bayesian}. Additionally, we introduce a three-level function modified from the Franke function \citep{franke1979critical}. The explicit forms of these functions are available in Supplementary Materials \ref{app:functions}.

The data are generated by evaluating these functions at input locations obtained from the nested space-filling design introduced by \cite{le2014recursive} with sample sizes $\{n_l\}^L_{l=1}$. 
The sample sizes and input dimension for each example are outlined in Table \ref{tab:simulationstudy}. To examine the prediction performance, $n_{\rm{test}}=1000$ random test input locations are generated from the same input space. We evaluate the prediction performance based on two criteria: the root-mean-square error (RMSE) and continuous rank probability score (CRPS) \citep{gneiting2007strictly}, which are defined in Supplementary Materials \ref{supp:RMSE}.
Note that CRPS serves as a performance metric for the \textit{posterior predictive distribution} of a scalar observation. Lower values for the RMSE and CRPS indicate better model accuracy. 
Additionally, we assess the computational efficiency by comparing the computation time.

\begin{figure}[!t]
\begin{center}
\includegraphics[width=0.9\textwidth]{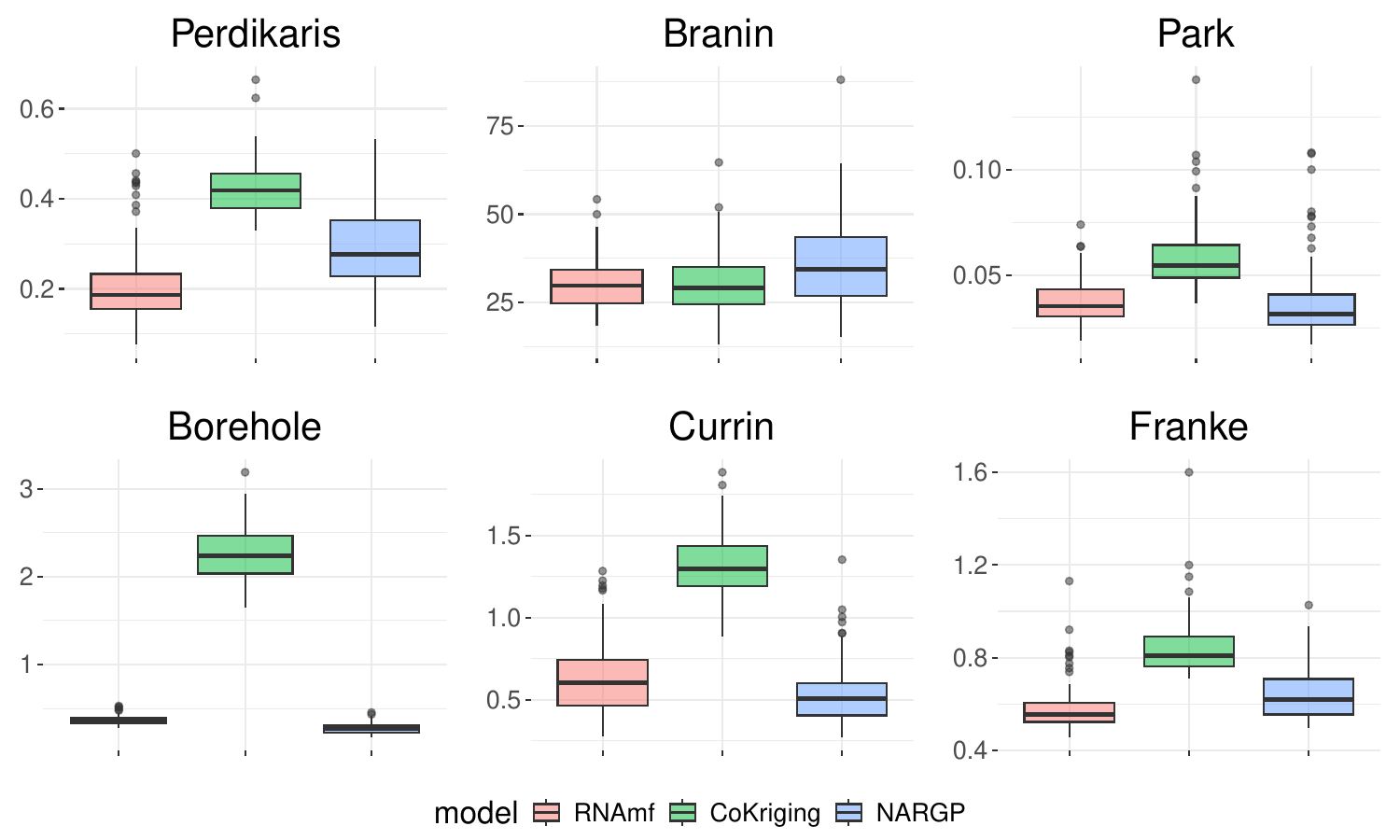} 
\end{center}
\caption{RMSEs of six synthetic examples across 100 repetitions.}
\label{fig:num_RMSE}
\end{figure}

\begin{figure}[!ht]
\begin{center}
\includegraphics[width=0.9\textwidth]{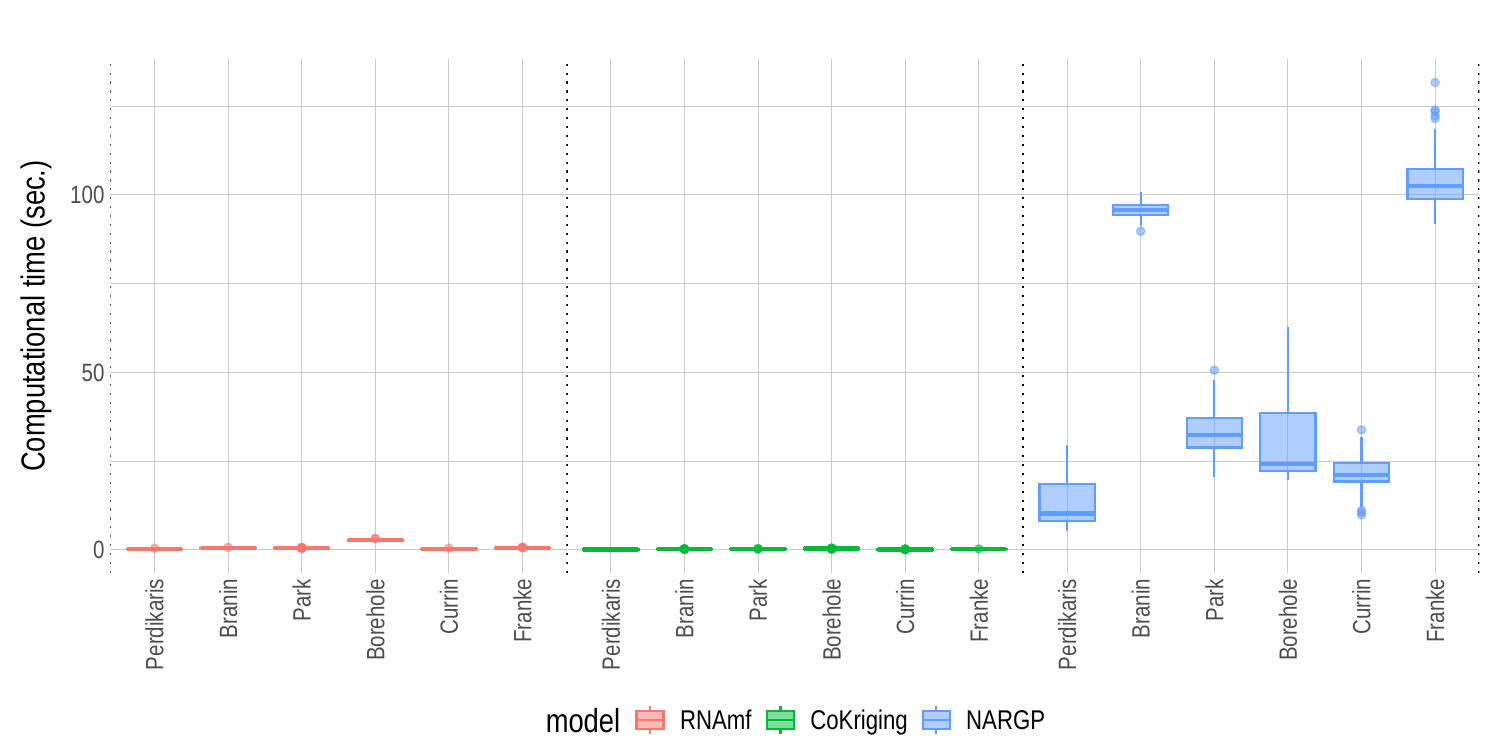} 
\end{center}
\caption{Computational time of six synthetic functions across 100 repetitions.}
\label{eq:num_computation}
\end{figure}

Figures \ref{fig:num_RMSE} and \ref{fig:num_CRPS} respectively show the results of RMSE and CRPS metrics across 100 repetitions, each employing a different random nested design for the training input locations. The proposed \texttt{RNAmf} consistently outperforms \texttt{CoKriging} by both metrics, particularly for examples exhibiting nonlinear relationships between simulators, such as the Perdikaris, Borehole, Currin, and Franke functions. For instances where simulators follow a  linear (or nearly linear) auto-regressive model, like the Brainin and Park functions, the proposed \texttt{RNAmf} remains competitive with \texttt{CoKriging}, which is designed to excel in such scenarios. This highlights the flexibility of our approach, enabled by the GP prior for modeling relationships. On the other hand, \texttt{NARGP}, another approach modeling nonlinear relationships, outperforms \texttt{CoKriging} in most of the examples and is competitive with \texttt{RNAmf}, except in the Perdikaris and Franke examples, where \texttt{RNAmf} exhibits superior performance. However, it comes with significantly higher computational costs, as shown in Figure \ref{eq:num_computation}, due to its expensive MC approximation, being roughly fifty times slower than both \texttt{RNAmf} and \texttt{CoKriging} on average. Notably, in scenarios involving three fidelities, including the Brainin and Franke examples, the computational time for \texttt{NARGP} exceeds that of \texttt{RNAmf} by more than 150 times. This shows that \texttt{NARGP} can suffer from intensive computation as the number of fidelity levels increases, while our method remains competitive in this regard. In summary, the performance across these synthetic examples underscores the capability of the proposed method in providing an accurate emulator at a reasonable computational time. 



\subsection{Active learning performance}\label{sec:ALperformance}
With the accurate RNA emulator in place, we now investigate on the performance of AL strategies for the emulator using the proposed criteria. We compare with two existing methods: \texttt{CoKriging-CV}, a cokriging-based sequential design utilizing cross-validation techniques \citep{le2015cokriging}, and \texttt{MR-SUR}, a sequential design maximizing the rate of stepwise uncertainty reduction using the KO model \citep{stroh2022sequential}. 
As for implementing \texttt{CoKriging-CV}, we utilized the code provided in the Supplementary Materials of \cite{le2015cokriging}.
Notably, both of these methods employed the (linear) autoregressive model as in \eqref{eq:autoregressive} in their implementations. To maintain a consistent comparison, we use the one-dimensional Perdikaris function (nonlinear) and the 4-dimensional Park function (linear autoregressive) in Section \ref{sec:emulationperformance}, to illustrate the performance of these methods. 

In this experiment, we suppose that the simulation costs associated with the low- and high-fidelity simulators are $C_1=1$ and $C_2=3$, respectively. The initial data is established similar to Section \ref{sec:emulationperformance}, with sample sizes specified in Table \ref{tab:simulationstudy}. We consider a total simulation budget of $C_{\rm{total}}=80$ for the Perdikaris function and $C_{\rm{total}}=130$ for the Park function. For ALC and ALMC acquisitions which involve the computation of the average reduction in variance as in \eqref{eq:averagevariancereduction}, 1000 and 100 uniform samples are respectively generated from the input space to approximate the integral and impute the future outputs.

\begin{figure}[ht!]
\begin{center}
\includegraphics[width=0.85\textwidth]{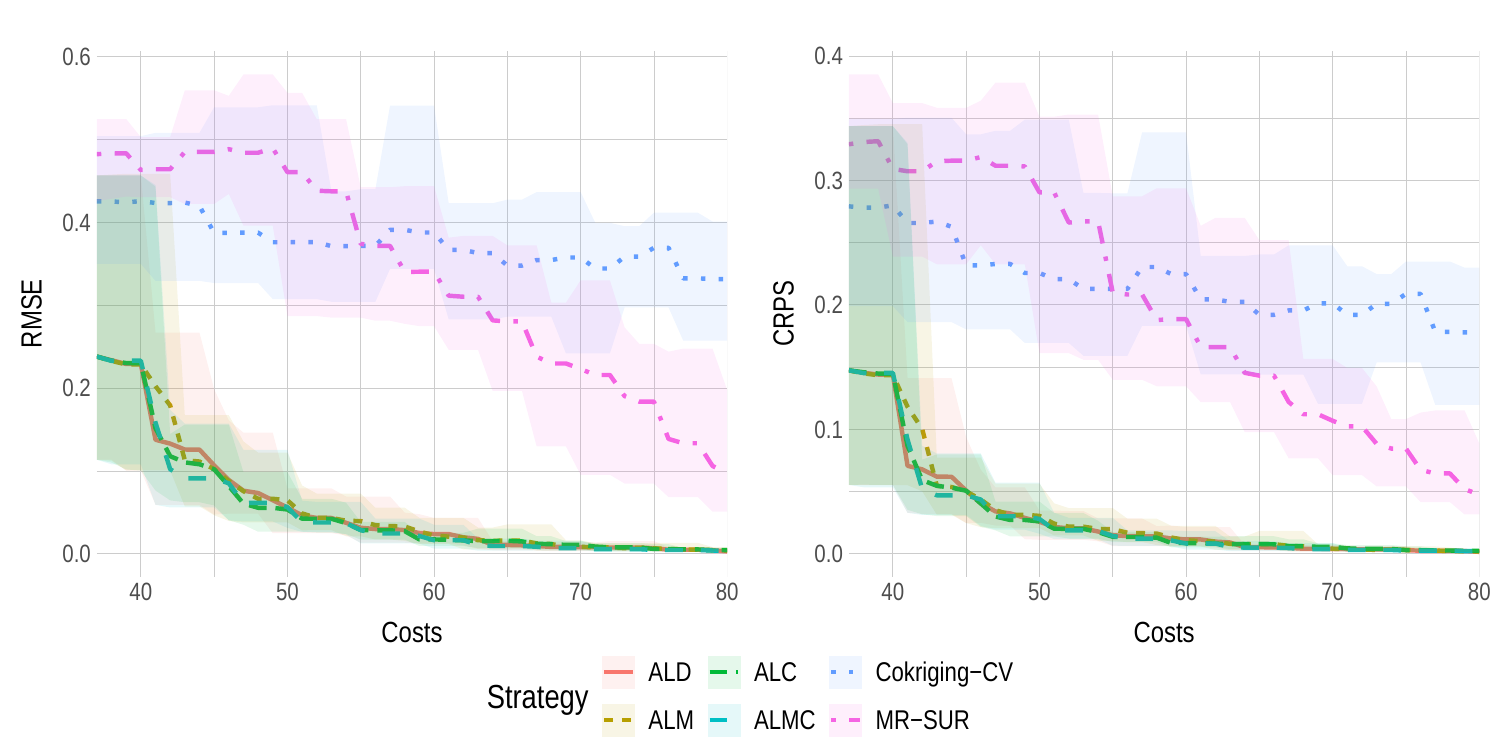} 
\end{center}
\caption{RMSE and CRPS for the Perdikaris function with respect to the simulation cost. Solid lines represent the average over 10 repetitions and shaded regions represent the ranges.}
\label{fig:AC_perdikaris}
\end{figure}

\begin{figure}[ht!]
\begin{center}
\includegraphics[width=0.8\textwidth]{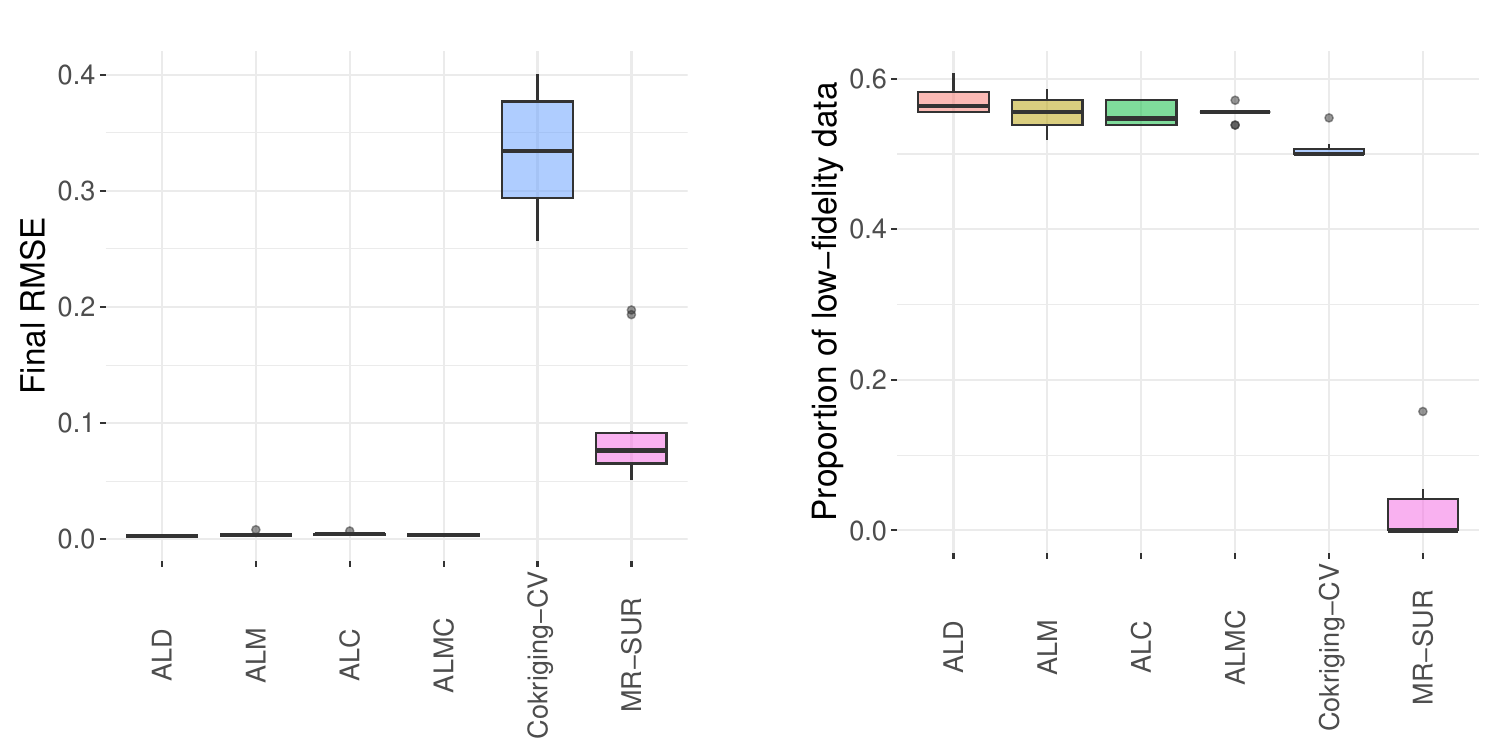} 
\end{center}
\caption{Final RMSE (left) and proportion of AL acquisitions choosing low-fidelity data (right) for the Perdikaris function. Boxplots indicate spread over 10 repetitions.}
\label{fig:AC_perdikaris_proportion}
\end{figure}

Figure \ref{fig:AC_perdikaris} shows the results of RMSE and CRPS metrics for the Perdikaris function, with respect to the total simulation costs accrued after each sample selection. The left panel of Figure \ref{fig:AC_perdikaris_proportion} displays a boxplot depicting the final RMSEs after reaching the total simulation budget across the 10 repetitions. The results show that the proposed AL methods dramatically outperform the two competing methods, \texttt{CoKriging-CV} and \texttt{MR-SUR}, in terms of both accuracy and stability, considering the same costs. As the cost increases, \texttt{MR-SUR} begins to close the gap, while \texttt{CoKriging-CV} lags behind the other methods. Among the four proposed AL strategies, the distinctions are minimal. 
As noted in Section \ref{sec:activelearning}, ALC acquisitions involve intricate numerical integration approximations and data imputation, taking approximately 400 seconds for each acquisition in this example. In contrast, ALD, ALM and ALMC are significantly more computationally efficient due to the closed-form nature of the criteria, requiring only around 1, 1, and 10 seconds per acquisition, respectively.

From the right panel of Figure \ref{fig:AC_perdikaris_proportion}, it can be seen that the proposed AL methods tend to select low-fidelity simulators more frequently than the other two comparative methods, notably \texttt{MR-SUR}, which consistently chooses samples exclusively from the high-fidelity simulator. This suggests that the proposed RNA model can effectively infer the high-fidelity simulation using primarily low-fidelity data for the nonlinear Perdikaris function, while the other two KO-based methods (\texttt{CoKriging-CV} and \texttt{MR-SUR}) require more high-fidelity data to reduce the uncertainty.

Figures \ref{fig:AC_park} and \ref{fig:AC_park_proportion} present the results for the Park function. As expected, the distinctions between these strategies are not as significant because the function aligns more closely with the KO model (linear autoregressive). Nonetheless, our proposed AL strategies still exhibit better average performance. At the final cost budget of $C_{\rm{total}}=130$, ALM and ALMC perform the best, collecting a larger portion of high-fidelity data, as indicated in Figure \ref{fig:AC_park_proportion}. In contrast, the KO-based strategies collect more low-fidelity data, which is again expected because KO-based models are efficient at leveraging low-fidelity data to infer the high-fidelity simulator. In these scenarios, our strategies efficiently prioritize the selection of high-fidelity data to minimize uncertainty, resulting in superior prediction accuracy at the same cost.



\section{Thermal Stress Analysis of Jet Engine Turbine Blade}\label{sec:realdata}
We leverage our proposed method for a real application involving the analysis of thermal stress in a jet turbine engine blade under steady-state operating conditions. The turbine blade, which forms part of the jet engine, is constructed from nickel alloys capable of withstanding extremely high temperatures. It is crucial for the blade's design to ensure that it can endure stress and deformations while avoiding mechanical failure and friction between the blade tip and the turbine casing. Refer to \cite{carter2005common}, \cite{wright2006enhanced}, and \cite{sung2022stacking,sung2023mcgp} for more details.

This problem can be treated as a static structural model and can be solved numerically using finite element methods. There are two input variables denoted as $x_1$ and  $x_2$, which represent the pressure load on the pressure and suction sides of the blade, both of which fall within the range of 0.25 to 0.75 MPa, i.e., $\mathbf{x}=(x_1,x_2)\in\Omega=[0.25,0.75]^2$. The response of interest is the maximum value over the thermal stress profile, which is a critical parameter used to assess the structural stability of the turbine blade. We perform finite element simulations using the \textsf{Partial Differential Equation Toolbox} in \textsf{MATLAB} \citep{MATLAB:R2021b}.

The simulations are conducted at two fidelity levels, each using different mesh densities for finite element methods. A denser mesh provides higher fidelity and more accurate results but demands greater computational resources. Conversely, a coarser mesh sacrifices some accuracy for reduced computational cost. Figure \ref{fig:turbine} demonstrates the turbine blade structure and thermal stress profiles obtained at these two fidelity levels for the input location $\mathbf{x}=(0.5,0.45)$.

We perform the finite element simulations with sample sizes of $n_1=20$ and $n_2=10$ to examine the emulation performance. Similar to Section \ref{sec:emulationperformance}, we use the nested space-filling design of \cite{le2014recursive} to generate the input locations of the computer experiments. We record the simulation time of the finite element simulations, which are respectively $C_1=2.25$ and $C_2=6.85$ (seconds) and will be used later for comparing AL strategies. To examine the performance, we conduct the high-fidelity simulations (i.e. $f_2(\mathbf{x})$) at the test input locations of size $n_{\text{test}}=100$ generated from a set of Latin hypercube samples from the same design space. The experiment is repeated 10 times, each time considering different nested space-filling designs for the training input locations.

\begin{figure}[ht!]
\begin{center}
\includegraphics[width=0.9\textwidth]{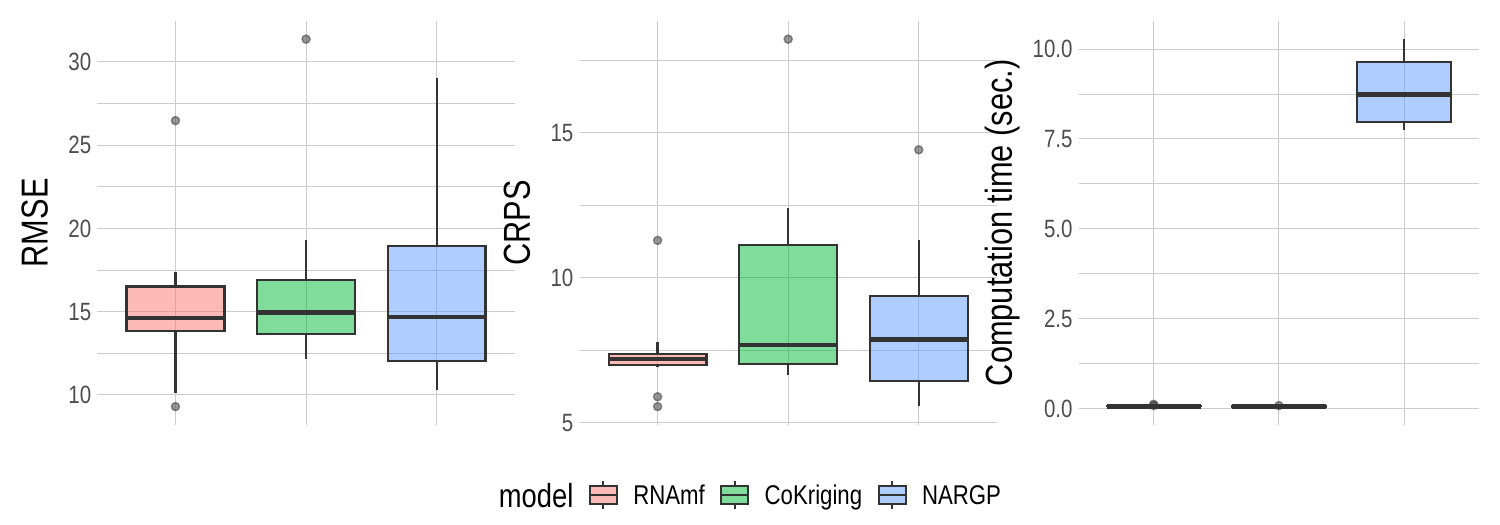} 
\end{center}
\caption{RMSE, CRPS, and computation time across 10 repetitions in the turbine blade application.}
\label{fig:emulationblade}
\end{figure}

Figure \ref{fig:emulationblade} presents a comparison of emulation performance with the other two competing methods, \texttt{CoKriging} and \texttt{NARGP}. Our proposed method, \texttt{RNAmf}, outperforms the other two methods in terms of CRPS and is comparable in terms of RMSE. While  \texttt{NARGP} delivers competitive prediction performance, it comes at a significantly higher computational cost compared to \texttt{RNAmf}.

\begin{figure}[ht!]
\begin{center}
\includegraphics[width=\textwidth]{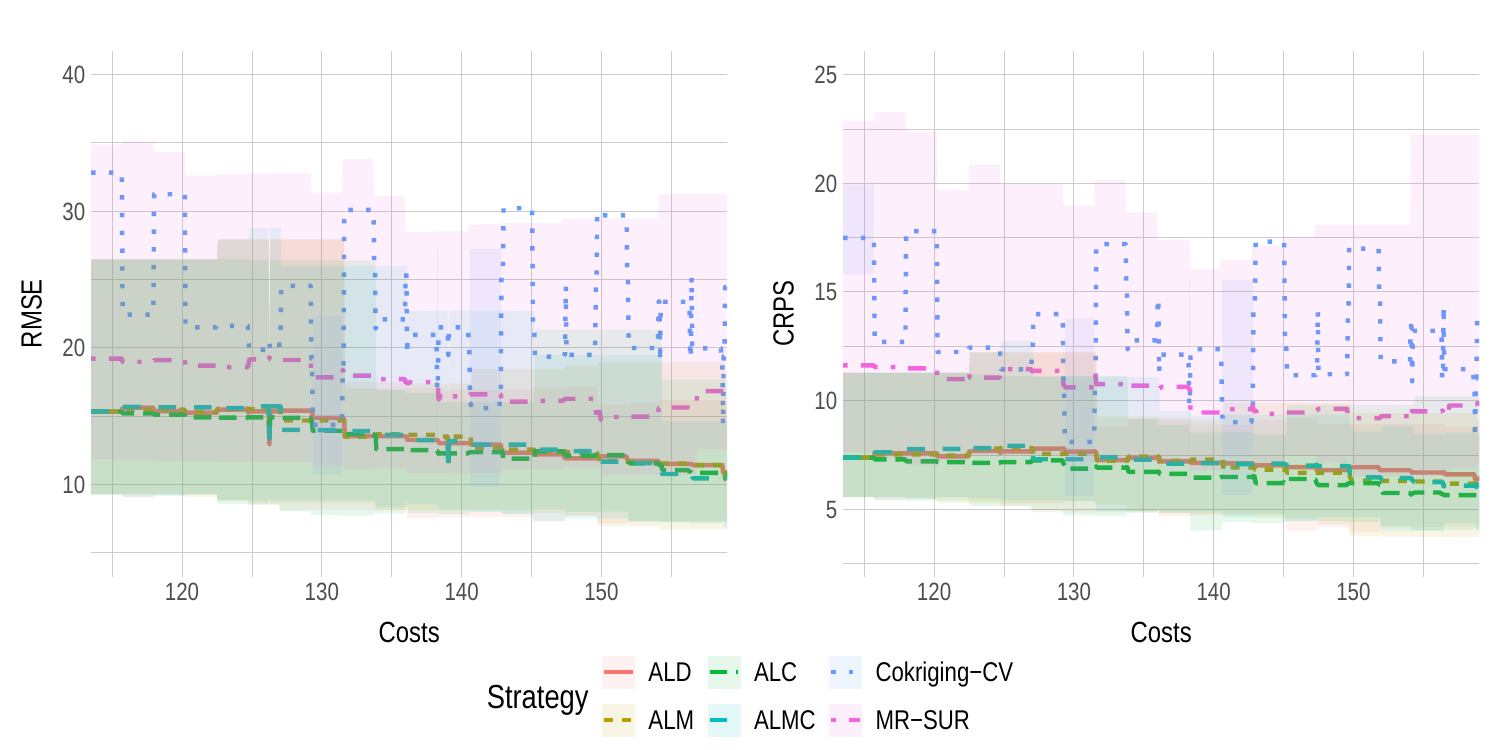} 
\end{center}
\caption{RMSE and CRPS for the turbine blade application with respect to the cost. Solid lines represent the average over 10 repetitions and shaded regions represent the ranges.}
\label{fig:AC_blade}
\end{figure}

\begin{figure}[ht!]
\begin{center}
\includegraphics[width=0.9\textwidth]{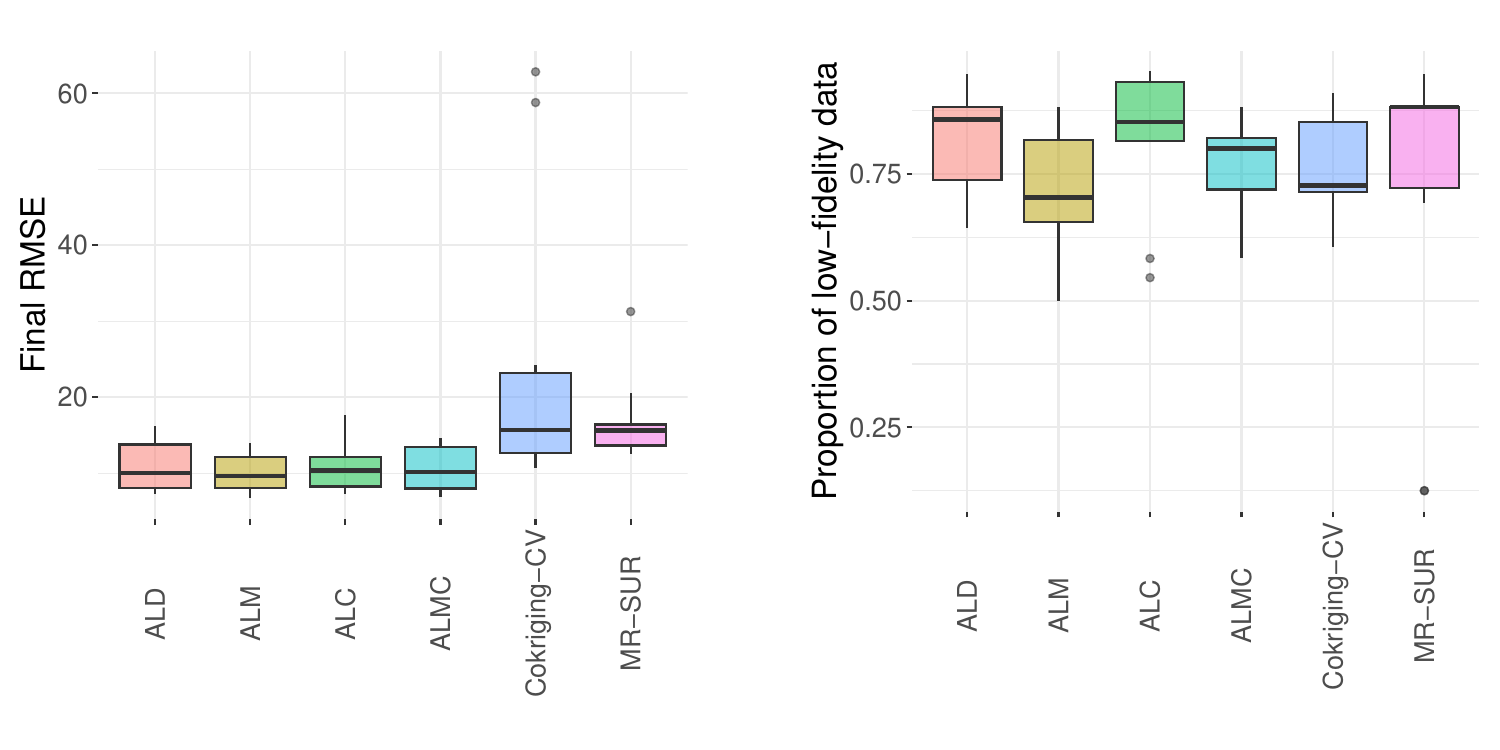} 
\end{center}
\caption{Final RMSE (left) and proportion of AL acquisitions choosing low-fidelity data (right) for the turbine blade application. Boxplots indicate spread over 10 repetitions.}
\label{fig:AC_blade_proportion}
\end{figure}

Figures \ref{fig:AC_blade} and \ref{fig:AC_blade_proportion} present a comparison of the AL strategies with a fixed cost budget of $C_{\rm{total}}=160$ seconds. The right panel of Figure \ref{fig:AC_blade_proportion} reveals that these strategies collect a similar number of low-fidelity data points. Notably, \texttt{CoKriging-CV} exhibits significant variability across the 10 repetitions, so we have removed the shaded region and only show the average, indicating that it yields poorer prediction performance compared to the other strategies. Another KO-based strategy, \texttt{MR-SUR}, performs better but still falls short of our proposed AL strategies at any given simulation cost. Conversely, our proposed AL strategies demonstrate effective results and outperform the others. This is evident from  RMSE and CRPS values exhibiting a leveling-off trend, with final results around 10 and 5, respectively, compared to the initial designs yielding both metrics averaging around 15 and 7. Among the AL strategies, the performance of the four strategies does not show significant differences at the final cost budget.

\section{Conclusion}\label{sec:conclusion}

Multi-fidelity computer experiments have become an essential tool in simulating complex scientific problems. This paper introduces a new emulator tailored for multi-fidelity simulations, which proves effective in producing accurate, efficient predictions for high-fidelity simulations, especially when dealing with nonlinear relationships between simulators. Building upon this new emulator, we present four AL strategies designed to select optimal input locations and fidelity levels to augment data, thereby enhancing
emulation performance. 

With the RNA emulator's success, it is worthwhile to explore emulators and AL strategies built upon similar principles for addressing multi-fidelity problems with \textit{tunable} fidelity parameters, such as mesh density \citep{picheny2013quantile,tuo2014surrogate}. Designing experiments for such scenarios presents intriguing challenges, as shown in recent studies (see, e.g., \citealp{shaowu2023design, sung2022stacking}). Furthermore, considering the increasing prevalence of \textit{stochastic computer models} \citep{baker2022analyzing}, extending the proposed RNA emulator to accommodate noisy data would significantly enhance its relevance in real-world applications. While this article assumes noise-free data, introducing noise into the model is a feasible endeavor, a task we leave for our future research. 


\vspace{0.5cm}
\noindent\textbf{Supplemental Materials}
Additional supporting materials can be found in Supplemental Materials, including the closed-form posterior mean and variance under a Mat\'ern kernel, the proof of Proposition \ref{prop:closedform}, and the supporting tables and figures for Sections \ref{sec:numericstudies} and \ref{sec:realdata}. The \textsf{R} code and package for reproducing the results in Sections
\ref{sec:numericstudies} and \ref{sec:realdata}  are also provided.

\bibliography{ref}

\newpage
\setcounter{page}{1}
\bigskip
\bigskip
\bigskip
\begin{center}
{\Large\bf Supplementary Materials for ``Active Learning for a Recursive Non-Additive Emulator for Multi-Fidelity Computer Experiments''}
\end{center}
\medskip

\setcounter{section}{0}
\setcounter{equation}{0}
\def\theequation{S\arabic{section}.\arabic{equation}}
\def\thesection{S\arabic{section}}
\def\thefigure{S\arabic{figure}}
\def\thetable{S\arabic{table}}

\section{Mat\'ern kernel functions}\label{app:matern}
Here are the Matérn kernels with smoothness parameters of 1.5 and 2.5, both of which are widely used and come with simpler expressions as follows: 
$$K_1(\mathbf{x}, \mathbf{x}')=\prod^d_{j=1}\psi(x_j,x'_j;\theta_{1j})$$
and
$$K_l(\mathbf{z}, \mathbf{z}')=\psi(y,y';\theta_{ly})\prod^d_{j=1}\psi(x_j,x'_j;\theta_{lj})$$
with
$$\psi(x,x';\theta) =\left( 1+\frac{\sqrt{3}|x- x'|}{\theta} \right) \exp \left( -\frac{\sqrt{3}|x- x'|}{\theta} \right)$$ for smoothness parameter of 1.5
and 
$$\psi(x, x';\theta) =  \left( 1+\frac{\sqrt{5}|x-x'|}{\theta} +\frac{5(x-x')^2}{3\theta^2} \right) \exp \left( -\frac{\sqrt{5}|x-x'|}{\theta} \right)$$ 
for smoothness parameter of  2.5.

\section{Proof of Proposition \ref{prop:closedform}}\label{supp:proof3.1}
For notational simplicity, we denote $\mathbf{Y}_l=\{\mathbf{y}_1,\ldots,\mathbf{y}_l\}$. Based on the squared exponential kernel, the posterior mean and variance at the input $\mathbf{x}$ can be derived as follows,
\begin{align*}
\mu_l^*(\mathbf{x})&=\mathbb{E}[f_l(\mathbf{x})|\mathbf{Y}_l] \\
&= \mathbb{E} [ \mathbb{E} [f_l(\mathbf{x}) | f_{l-1}(\mathbf{x}), \mathbf{Y}_l ] ] \\
&= \alpha_l + \mathbb{E} [ \mathbf{k}_l (\mathbf{x},f_{l-1}(\mathbf{x}))^T |\mathbf{Y}_l] \mathbf{K}^{-1}_l (\mathbf{y}_l - \alpha_l \mathbf{1}_{n_l}) \\
&= \alpha_l + \sum^{n_l}_{i=1} r_i \prod_{j=1}^d \exp\left( -\frac{(x_{j}-x^{[l]}_{ij})^2}{\theta_{lj}} \right)  \mathbb{E} \left[ \exp{\left\{ -\frac{(y_i^{[l-1]}-f_{l-1}(\mathbf{x}))^2}{\theta_{ly}} \right\} } \Big|\mathbf{Y}_l \right] \\
&= \alpha_l + \sum^{n_l}_{i=1} r_i \prod_{j=1}^d \exp\left( -\frac{(x_{j}-x^{[l]}_{ij})^2}{\theta_{lj}} \right)  \frac{1}{\sqrt{1+2\frac{
\sigma^{*2}_{l-1}(\mathbf{x}) }{\theta_{ly}}}}  \exp{\left( -\frac{(y_i^{[l-1]}-\mu^*_{l-1}(\mathbf{x}))^2}{\theta_{ly}+2\sigma^{*2}_{l-1}(\mathbf{x})} \right)} ,\\
\text{and}\\
\sigma_l^{*2}(\mathbf{x})&=\mathbb{V}[f_l(\mathbf{x})|\mathbf{Y}_l] \\
&= \mathbb{V} \left[ \mathbb{E} [f_l(\mathbf{x}) | f_{l-1}(\mathbf{x}), \mathbf{Y}_l ] \right] + \mathbb{E} \left[\mathbb{V} [f_l(\mathbf{x}) | f_{l-1}(\mathbf{x}), \mathbf{Y}_l ] \right] \\
&= \mathbb{E} \left[ \left\{ \mathbb{E} \left[f_l(\mathbf{x}) | f_{l-1}(\mathbf{x}), \mathbf{Y}_l \right] \right\}^2 \right] - \mu^{*}_l(\mathbf{x})^2 + \mathbb{E} \left[\mathbb{V} \left[f_l(\mathbf{x}) | f_{l-1}(\mathbf{x}), \mathbf{Y}_l \right] \right] \\
&= \mathbb{E}\left[ \left\{\alpha_l +  \mathbf{k}_l (\mathbf{x},f_{l-1}(\mathbf{x}))^T   \mathbf{K}^{-1}_l  (\mathbf{y}_l - \alpha_l \mathbf{1}_{n_l}) \right\}^2 |\mathbf{Y}_l\right] -\mu^{*}_l(\mathbf{x})^2  \\
&\quad\quad\quad\quad\quad\quad\quad\quad\quad+\mathbb{E} \left[ \tau_l^2 \left\{ 1 - \mathbf{k}_l (\mathbf{x},f_{l-1}(\mathbf{x}))^T  \mathbf{K}^{-1}_l  \mathbf{k}_l(\mathbf{x},f_{l-1}(\mathbf{x})) \right\} |\mathbf{Y}_l \right] \\
&= \alpha_l^2 + 2\alpha_l  (\mu^*_l(\mathbf{x}) - \alpha_l) + \mathbb{E}\left[ \left\{ \mathbf{k}_l (\mathbf{x},f_{l-1}(\mathbf{x}))^T  \mathbf{K}^{-1}_l  (\mathbf{y}_l - \alpha_l \mathbf{1}_{n_l}) \right\}^2 |\mathbf{Y}_l\right] \\
&\quad\quad\quad\quad\quad\quad\quad\quad\quad-\mu^{*}_l(\mathbf{x})^2 + \tau^2_l - \tau^2_l \mathbb{E} \left[ \left\{ \mathbf{k}_l (\mathbf{x},f_{l-1}(\mathbf{x}))^T  \mathbf{K}^{-1}_l  \mathbf{k}_l(\mathbf{x},f_{l-1}(\mathbf{x}))  \right\}  |\mathbf{Y}_l\right]\\
&= \tau^2_l - (\mu^*_l(\mathbf{x})-\alpha_l)^2 +\left( \sum_{i,k=1}^{n_l} \zeta_{ik} \left(r_i r_k - \tau^2_l (\mathbf{K}_l^{-1})_{ik} \right) \prod_{j=1}^d \exp{ \left(-\frac{(x_{j}-{x}^{[l]}_{ij})^2+(x_{j}-{x}^{[l]}_{kj})^2}{\theta_{lj}}\right)}  \right), 
\end{align*}
where 
\begin{align*}
r_i &= (\mathbf{K}^{-1}_l (\mathbf{y}_l - \alpha_l \mathbf{1}_{n_l}))_i, \\
\zeta_{ik} &= \frac{1}{\sqrt{1+4\frac{\sigma^{*2}_{l-1}(\mathbf{x})}{\theta_{ly}}}}  \exp{\left( -\frac{(\frac{y_i^{[l-1]}+y_k^{[l-1]}}{2}-\mu^*_{l-1}(\mathbf{x}))^2}{\frac{\theta_{ly}}{2}+2\sigma^{*2}_{l-1}(\mathbf{x})} -\frac{(y_i^{[l-1]}-y_k^{[l-1]})^2}{2\theta_{ly}} \right)}. 
\end{align*}

\section{Posterior mean and variance under Mat\'ern kernel}\label{supp:maternposterior}

This section is developed along the line of \cite{suppming2021}. The posterior mean and variance at the input $\mathbf{x}$ can be derived as follows,
\begin{align*}
\mu^*_l (\mathbf{x})&=\mathbb{E}[f_l(\mathbf{x})|\mathbf{Y}_l] \\
&= \alpha_l + \mathbb{E} [ \mathbf{k}_l (\mathbf{x},f_{l-1}(\mathbf{x}))^T |\mathbf{Y}_l] \mathbf{K}^{-1}_l (\mathbf{y}_l - \alpha_l \mathbf{1}_{n_l}) \\
&= \alpha_l + \sum^{n_l}_{i=1} r_i \xi_i \prod^d_{j=1}\psi(x_j,x^{[l]}_{ij};\theta_{lj})
, \\
\sigma_l^{*2}(\mathbf{x})&=\mathbb{V}[f_l(\mathbf{x})|\mathbf{Y}_l] \\
&= \alpha_l^2 + 2\alpha_l  (\mu^*_l(\mathbf{x}) - \alpha_l) + \mathbb{E}\left[ \left\{ \mathbf{k}_l (\mathbf{x},f_{l-1}(\mathbf{x}))^T  \mathbf{K}^{-1}_l  (\mathbf{y}_l - \alpha_l \mathbf{1}_{n_l}) \right\}^2 |\mathbf{Y}_l\right] \\
&\quad\quad\quad\quad\quad\quad\quad\quad\quad-\mu^{*}_l(\mathbf{x})^2 + \tau^2_l - \tau^2_l \mathbb{E} \left[ \left\{ \mathbf{k}_l (\mathbf{x},f_{l-1}(\mathbf{x}))^T  \mathbf{K}^{-1}_l  \mathbf{k}_l(\mathbf{x},f_{l-1}(\mathbf{x}))  \right\}  |\mathbf{Y}_l\right]\\
&= \tau^2_l - (\mu^*_l(\mathbf{x})-\alpha_l)^2 +\left( \sum_{i,k=1}^{n_l} \zeta_{ik} \left(r_i r_k - \tau^2_l (\mathbf{K}_l^{-1})_{ik} \right) \prod_{j=1}^d \psi(x_j,x^{[l]}_{ij};\theta_{lj}) \psi(x_j,x^{[l]}_{kj};\theta_{lj}) \right)
\end{align*}
where $\psi$ is defined in Section \ref{app:matern}, 
$(r_1, \ldots, r_{n_l})^T = \mathbf{K}^{-1}_l (\mathbf{y}_l - \alpha_l \mathbf{1}_{n_l}),\xi_i = \mathbb{E} \left[ \psi(f_{l-1}(\mathbf{x}), y_i^{[l-1]} ; \theta_{ly}) \Big|\mathbf{Y}_l \right]$, and $\zeta_{ik} = \mathbb{E} \left[ \psi(f_{l-1}(\mathbf{x}), y_i^{[l-1]} ; \theta_{ly}) \psi(f_{l-1}(\mathbf{x}), y_k^{[l-1]} ; \theta_{ly}) \Big|\mathbf{Y}_l \right]$. The closed-form expressions of $\xi_i$ and $\zeta_{ik}$ are provided in the following subsections.

\subsection{Mat\'ern-1.5 kernel}

For Mat\'ern kernel with the smoothness parameter of 1.5, $\xi_i$ and $\zeta_{ik}$ are provided as follows,
\begin{align*}
\xi_i &= \exp \left( \frac{3 \sigma^{*2}_{l-1}(\mathbf{x}) + 2\sqrt{3} \theta_{ly} (y_i^{[l-1]} - \mu^*_{l-1}(\mathbf{x}) ) }{2 \theta_{ly}^2} \right) \times \left[ E_1' \Lambda_{11} \Phi \left( \frac{\mu^*_{l-1}(\mathbf{x}) - y_i^{[l-1]} - \frac{\sqrt{3}\sigma^{*2}_{l-1}(\mathbf{x})}{\theta_{ly}}}{\sigma^{*}_{l-1}(\mathbf{x})}  \right) \right. \\
&+ \left. E_1' \Lambda_{12} \frac{\sigma^{*}_{l-1}(\mathbf{x})}{\sqrt{2\pi}} \exp \left( -\frac{ \left(y_i^{[l-1]} - \mu^*_{l-1}(\mathbf{x}) + \frac{\sqrt{3}\sigma^{*2}_{l-1}(\mathbf{x})}{\theta_{ly}} \right)^2}{2\sigma^{*2}_{l-1}(\mathbf{x})} \right) \right] \\
&+ \exp \left( \frac{3 \sigma^{*2}_{l-1}(\mathbf{x}) - 2\sqrt{3} \theta_{ly} (y_i^{[l-1]} - \mu^*_{l-1}(\mathbf{x}) ) }{2 \theta_{ly}^2} \right) \times \left[ E_2' \Lambda_{21}  \Phi \left( \frac{-\mu^*_{l-1}(\mathbf{x}) + y_i^{[l-1]} - \frac{\sqrt{3}\sigma^{*2}_{l-1}(\mathbf{x})}{\theta_{ly}}}{\sigma^{*}_{l-1}(\mathbf{x})}  \right) \right. \\
&+ \left. E_2' \Lambda_{12}  \cdot \frac{\sigma^{*}_{l-1}(\mathbf{x})}{\sqrt{2\pi}} \exp \left( -\frac{ \left( y_i^{[l-1]} - \mu^*_{l-1}(\mathbf{x}) - \frac{\sqrt{3}\sigma^{*2}_{l-1}(\mathbf{x})}{\theta_{ly}} \right)^2 }{2\sigma^{*2}_{l-1}(\mathbf{x})} \right) \right], \\
\zeta_{ik} &= \exp \left\{ \frac{6\sigma^{*2}_{l-1}(\mathbf{x}) + \sqrt{3} \theta_{ly} \left( y_i^{[l-1]}+y_k^{[l-1]} -2\mu^*_{l-1}(\mathbf{x}) \right)}{\theta_{ly}^2}  \right\} \\
&\times \left[ E_{3}' \Lambda_{31}
\Phi \left\{ \frac{ \left( \mu^*_{l-1}(\mathbf{x}) -y_k^{[l-1]} - 2\sqrt{3} \frac{\sigma^{*2}_{l-1}(\mathbf{x})}{\theta_{ly} }  \right) }{\sigma^{*}_{l-1}(\mathbf{x})}  \right. \right\} \\
&+ E_{3}' \Lambda_{32} \left. \frac{\sigma^{*}_{l-1}(\mathbf{x})}{\sqrt{2\pi}} \exp \left(-\frac{\left(y_k^{[l-1]}-\mu^*_{l-1}(\mathbf{x})+ 2\sqrt{3} \frac{\sigma^{*2}_{l-1}(\mathbf{x})}{\theta_{ly} }  \right)^2}{2\sigma^{*2}_{l-1}(\mathbf{x})}  \right)
\right] + \exp \left\{ - \frac{\sqrt{3} \left( y_k^{[l-1]}-y_i^{[l-1]} \right)}{\theta_{ly}}  \right\} \\
&\times \left[ E_{4}' \Lambda_{41} \left( \Phi \left\{ \frac{ y_k^{[l-1]} - \mu^*_{l-1}(\mathbf{x}) }{\sigma^{*}_{l-1}(\mathbf{x})}  \right\} -  \Phi \left\{ \frac{ y_i^{[l-1]} - \mu^*_{l-1}(\mathbf{x}) }{\sigma^{*}_{l-1}(\mathbf{x})}  \right\} \right) \right. \\
&+ E_{4}' \Lambda_{42} \frac{\sigma^{*}_{l-1}(\mathbf{x})}{\sqrt{2\pi}} \exp \left(-\frac{\left(y_i^{[l-1]}-\mu^*_{l-1}(\mathbf{x}) \right)^2}{2\sigma^{*2}_{l-1}(\mathbf{x})} \right)   \left. - E_{4}' \Lambda_{43} \frac{\sigma^{*}_{l-1}(\mathbf{x})}{\sqrt{2\pi}} \exp \left(-\frac{\left(y_k^{[l-1]}-\mu^*_{l-1}(\mathbf{x}) \right)^2}{2\sigma^{*2}_{l-1}(\mathbf{x})} \right) \right] \\
&+ \exp \left\{ \frac{6\sigma^{*2}_{l-1}(\mathbf{x}) - \sqrt{3} \theta_{ly} \left( y_i^{[l-1]}+y_k^{[l-1]} -2\mu^*_{l-1}(\mathbf{x}) \right)}{\theta_{ly}^2}  \right\} \\
&\times \left[ E_{5}' \Lambda_{51}
\Phi \left\{ \frac{ \left( -\mu^*_{l-1}(\mathbf{x}) + y_k^{[l-1]} - 2\sqrt{3} \frac{\sigma^{*2}_{l-1}(\mathbf{x})}{\theta_{ly} }  \right) }{\sigma^{*}_{l-1}(\mathbf{x})}  \right\} \right. 
\end{align*}

\begin{align*}
&+ E_{5}' \Lambda_{52} \left. \frac{\sigma^{*}_{l-1}(\mathbf{x})}{\sqrt{2\pi}} \exp \left(-\frac{\left(y_i^{[l-1]}-\mu^*_{l-1}(\mathbf{x})- 2\sqrt{3} \frac{\sigma^{*2}_{l-1}(\mathbf{x})}{\theta_{ly} }  \right)^2}{2\sigma^{*2}_{l-1}(\mathbf{x})}  \right)
\right],\\
\Lambda_{11} &= 
\begin{pmatrix} 
1 \\ 
\mu^*_{l-1}(\mathbf{x}) - \frac{\sqrt{3}\sigma^{*2}_{l-1}(\mathbf{x})}{\theta_{ly}} 
\end{pmatrix},
\Lambda_{12} =
\begin{pmatrix} 
0 \\ 
1 
\end{pmatrix},
\Lambda_{21} = 
\begin{pmatrix} 
1 \\ 
- \mu^*_{l-1}(\mathbf{x}) - \frac{\sqrt{3}\sigma^{*2}_{l-1}(\mathbf{x})}{\theta_{ly}} 
\end{pmatrix},
\\
\Lambda_{31} &= 
\begin{pmatrix} 
1 \\ 
\mu^*_{l-1}(\mathbf{x}) - \frac{2\sqrt{3}\sigma^{*2}_{l-1}(\mathbf{x})}{\theta_{ly}} \\
\left( \mu^*_{l-1}(\mathbf{x}) - \frac{2\sqrt{3}\sigma^{*2}_{l-1}(\mathbf{x})}{\theta_{ly}} \right)^2 + \sigma^{*2}_{l-1}(\mathbf{x})
\end{pmatrix},
\Lambda_{32} =
\begin{pmatrix} 
0 \\ 
1 \\
\mu^*_{l-1}(\mathbf{x}) - \frac{2\sqrt{3}\sigma^{*2}_{l-1}(\mathbf{x})}{\theta_{ly}} + y_k^{[l-1]}
\end{pmatrix}, \\
\Lambda_{41} &= 
\begin{pmatrix} 
1 \\ 
\mu^*_{l-1}(\mathbf{x}) \\
\left( \mu^*_{l-1}(\mathbf{x}) \right)^2 + \sigma^{*2}_{l-1}(\mathbf{x})
\end{pmatrix},
\Lambda_{42} =
\begin{pmatrix} 
0 \\ 
1 \\
\mu^*_{l-1}(\mathbf{x}) + y_i^{[l-1]}
\end{pmatrix},\\
\Lambda_{43} &=
\begin{pmatrix} 
0 \\ 
1 \\
\mu^*_{l-1}(\mathbf{x}) + y_k^{[l-1]}
\end{pmatrix},
\Lambda_{51} = 
\begin{pmatrix} 
1 \\ 
-\mu^*_{l-1}(\mathbf{x}) - \frac{2\sqrt{3}\sigma^{*2}_{l-1}(\mathbf{x})}{\theta_{ly}} \\
\left( -\mu^*_{l-1}(\mathbf{x}) - \frac{2\sqrt{3}\sigma^{*2}_{l-1}(\mathbf{x})}{\theta_{ly}} \right)^2 + \sigma^{*2}_{l-1}(\mathbf{x})
\end{pmatrix},\\
\Lambda_{52} &=
\begin{pmatrix} 
0 \\ 
1 \\
-\mu^*_{l-1}(\mathbf{x}) - \frac{2\sqrt{3}\sigma^{*2}_{l-1}(\mathbf{x})}{\theta_{ly}} - y_i^{[l-1]}
\end{pmatrix},
E_{1} = \frac{1}{\theta_{ly}}
\begin{pmatrix} 
\theta_{ly} - \sqrt{3}y_i^{[l-1]} \\ 
\sqrt{3}
\end{pmatrix},\\
E_{2} &= \frac{1}{\theta_{ly}}
\begin{pmatrix} 
\theta_{ly} + \sqrt{3}y_i^{[l-1]} \\ 
\sqrt{3}
\end{pmatrix},
E_{3} = \frac{1}{\theta_{ly}^2}
\begin{pmatrix} 
\theta_{ly}^2 + 3y_i^{[l-1]}y_k^{[l-1]} - \sqrt{3}\theta_{ly}(y_i^{[l-1]}+y_k^{[l-1]}) \\ 
2\sqrt{3}\theta_{ly} - 3(y_i^{[l-1]}+y_k^{[l-1]})\\
3
\end{pmatrix},
\\
E_{4} &= \frac{1}{\theta_{ly}^2}
\begin{pmatrix} 
\theta_{ly}^2 - 3y_i^{[l-1]}y_k^{[l-1]} + \sqrt{3}\theta_{ly}(y_k^{[l-1]}-y_i^{[l-1]}) \\ 
3(y_i^{[l-1]}+y_k^{[l-1]})\\
-3
\end{pmatrix},
\\
E_{5} &= \frac{1}{\theta_{ly}^2}
\begin{pmatrix} 
\theta_{ly}^2 + 3y_i^{[l-1]}y_k^{[l-1]} + \sqrt{3}\theta_{ly}(y_i^{[l-1]}+y_k^{[l-1]}) \\ 
2\sqrt{3}\theta_{ly} + 3(y_i^{[l-1]}+y_k^{[l-1]})\\
3
\end{pmatrix},
\end{align*}
for $ y_i^{[l-1]} \leq y_k^{[l-1]}$. If $ y_i^{[l-1]} > y_k^{[l-1]}$, interchange $ y_i^{[l-1]}$ and $ y_k^{[l-1]}$. $\Phi$ is the cumulative distribution function of a standard normal distribution.

\subsection{Mat\'ern-2.5 kernel}
For Mat\'ern kernel with the smoothness parameter of 2.5, $\xi_i$ and $\zeta_{ik}$ are provided as follows, 
\begin{align*} 
\xi_i &= \exp \left( \frac{5 \sigma^{*2}_{l-1}(\mathbf{x}) + 2\sqrt{5} \theta_{ly} (y_i^{[l-1]} - \mu^*_{l-1}(\mathbf{x}) ) }{2 \theta_{ly}^2} \right) \times \left[ E_1' \Lambda_{11} \Phi \left( \frac{\mu^*_{l-1}(\mathbf{x}) - y_i^{[l-1]} - \frac{\sqrt{5}\sigma^{*2}_{l-1}(\mathbf{x})}{\theta_{ly}}}{\sigma^{*}_{l-1}(\mathbf{x})}  \right) \right. \\
&+ \left. E_1' \Lambda_{12} \frac{\sigma^{*}_{l-1}(\mathbf{x})}{\sqrt{2\pi}} \exp \left( -\frac{ \left( \mu^*_{l-1}(\mathbf{x}) - y_i^{[l-1]} - \frac{\sqrt{5}\sigma^{*2}_{l-1}(\mathbf{x})}{\theta_{ly}} \right)^2}{2\sigma^{*2}_{l-1}(\mathbf{x})} \right) \right] \\
&+ \exp \left( \frac{5 \sigma^{*2}_{l-1}(\mathbf{x}) - 2\sqrt{5} \theta_{ly} (y_i^{[l-1]} - \mu^*_{l-1}(\mathbf{x}) ) }{2 \theta_{ly}^2} \right) \times \left[ E_2' \Lambda_{21}  \Phi \left( \frac{-\mu^*_{l-1}(\mathbf{x}) + y_i^{[l-1]} - \frac{\sqrt{5}\sigma^{*2}_{l-1}(\mathbf{x})}{\theta_{ly}}}{\sigma^{*}_{l-1}(\mathbf{x})}  \right) \right. \\
&+ \left. E_2' \Lambda_{22}  \cdot \frac{\sigma^{*}_{l-1}(\mathbf{x})}{\sqrt{2\pi}} \exp \left( -\frac{ \left( - \mu^*_{l-1}(\mathbf{x}) + y_i^{[l-1]} - \frac{\sqrt{5}\sigma^{*2}_{l-1}(\mathbf{x})}{\theta_{ly}} \right)^2 }{2\sigma^{*2}_{l-1}(\mathbf{x})} \right) \right], 
\\
\zeta_{ik} &= \exp \left\{ \frac{10\sigma^{*2}_{l-1}(\mathbf{x}) + \sqrt{5} \theta_{ly} \left( y_i^{[l-1]}+y_k^{[l-1]} -2\mu^*_{l-1}(\mathbf{x}) \right)}{\theta_{ly}^2}  \right\} \\
&\times \left[ E_{3}' \Lambda_{31}
\Phi \left\{ \frac{ \left( \mu^*_{l-1}(\mathbf{x}) - y_k^{[l-1]} - 2\sqrt{5} \frac{\sigma^{*2}_{l-1}(\mathbf{x})}{\theta_{ly} }  \right) }{\sigma^{*}_{l-1}(\mathbf{x})}  \right. \right\} \\
&+ E_{3}' \Lambda_{32} \left. \frac{\sigma^{*}_{l-1}(\mathbf{x})}{\sqrt{2\pi}} \exp \left(-\frac{\left(\mu^*_{l-1}(\mathbf{x}) - y_k^{[l-1]} - 2\sqrt{5} \frac{\sigma^{*2}_{l-1}(\mathbf{x})}{\theta_{ly} }  \right)^2}{2\sigma^{*2}_{l-1}(\mathbf{x})}  \right)
\right] + \exp \left\{ - \frac{\sqrt{5} \left( y_k^{[l-1]}-y_i^{[l-1]} \right)}{\theta_{ly}}  \right\} \\
&\times \left[ E_{4}' \Lambda_{41} \left( \Phi \left\{ \frac{ y_k^{[l-1]} - \mu^*_{l-1}(\mathbf{x}) }{\sigma^{*}_{l-1}(\mathbf{x})}  \right\} -  \Phi \left\{ \frac{ y_i^{[l-1]} - \mu^*_{l-1}(\mathbf{x}) }{\sigma^{*}_{l-1}(\mathbf{x})}  \right\} \right) \right. \\
&+ E_{4}' \Lambda_{42} \frac{\sigma^{*}_{l-1}(\mathbf{x})}{\sqrt{2\pi}} \exp \left(-\frac{\left(y_i^{[l-1]}-\mu^*_{l-1}(\mathbf{x}) \right)^2}{2\sigma^{*2}_{l-1}(\mathbf{x})} \right)  \left. - E_{4}' \Lambda_{43} \frac{\sigma^{*}_{l-1}(\mathbf{x})}{\sqrt{2\pi}} \exp \left(-\frac{\left(y_k^{[l-1]}-\mu^*_{t-1}(\mathbf{x}) \right)^2}{2\sigma^{*2}_{l-1}(\mathbf{x})} \right) \right] 
\end{align*}

\begin{align*}
&+ \exp \left\{ \frac{10\sigma^{*2}_{l-1}(\mathbf{x}) - \sqrt{5} \theta_{ly} \left( y_i^{[l-1]}+y_k^{[l-1]} -2\mu^*_{l-1}(\mathbf{x}) \right)}{\theta_{ly}^2}  \right\} \\
& \times \left[ E_{5}' \Lambda_{51}
\Phi \left\{ \frac{ \left( -\mu^*_{l-1}(\mathbf{x}) + y_k^{[l-1]} - 2\sqrt{5} \frac{\sigma^{*2}_{l-1}(\mathbf{x})}{\theta_{ly} }  \right) }{\sigma^{*}_{l-1}(\mathbf{x})}  \right\} \right. \\
&+ E_{5}' \Lambda_{52} \left. \frac{\sigma^{*}_{l-1}(\mathbf{x})}{\sqrt{2\pi}} \exp \left(-\frac{\left(-\mu^*_{l-1}(\mathbf{x}) + y_i^{[l-1]} - 2\sqrt{5} \frac{\sigma^{*2}_{l-1}(\mathbf{x})}{\theta_{ly} }  \right)^2}{2\sigma^{*2}_{l-1}(\mathbf{x})}  \right)
\right],\\
\Lambda_{11} &= 
\begin{pmatrix} 
1 \\ 
\mu^*_{t-1}(\mathbf{x}) - \frac{\sqrt{5}\sigma^{*2}_{l-1}(\mathbf{x})}{\theta_{ly}} \\
\left( \mu^*_{l-1}(\mathbf{x}) - \frac{\sqrt{5}\sigma^{*2}_{l-1}(\mathbf{x})}{\theta_{ly}} \right)^2 + \sigma^{*2}_{l-1}(\mathbf{x})
\end{pmatrix},
\Lambda_{12} =
\begin{pmatrix} 
0 \\ 
1 \\
\mu^*_{l-1}(\mathbf{x}) - \frac{\sqrt{5}\sigma^{*2}_{l-1}(\mathbf{x})}{\theta_{ly}} + y_i^{[l-1]}
\end{pmatrix},\\
\Lambda_{21} &= 
\begin{pmatrix} 
1 \\ 
- \mu^*_{l-1}(\mathbf{x}) - \frac{\sqrt{5}\sigma^{*2}_{l-1}(\mathbf{x})}{\theta_{ly}} \\
\left( \mu^*_{l-1}(\mathbf{x}) + \frac{\sqrt{5}\sigma^{*2}_{l-1}(\mathbf{x})}{\theta_{ly}} \right)^2 + y_i^{[l-1]}
\end{pmatrix},
\Lambda_{22} =
\begin{pmatrix} 
0 \\ 
1 \\
-\mu^*_{l-1}(\mathbf{x}) - \frac{\sqrt{5}\sigma^{*2}_{l-1}(\mathbf{x})}{\theta_{ly}} - y_i^{[l-1]}
\end{pmatrix}, \\
\Lambda_{31} &= 
\begin{pmatrix} 
1 \\ 
\mu_c\\
\mu_c^2+\sigma^{*2}_{l-1}(\mathbf{x})\\
\mu_c \left( \mu_c^2+3\sigma^{*2}_{l-1}(\mathbf{x}) \right) \\
\mu_c^4 + 6\mu_c^2\sigma^{*2}_{l-1}(\mathbf{x}) + 3\sigma^{*4}_{l-1}(\mathbf{x})
\end{pmatrix}, \mu_c=\mu^*_{l-1}(\mathbf{x}) - 2\sqrt{5}\frac{\sigma^{*2}_{l-1}(\mathbf{x})}{\theta_{ly}},
\\
\Lambda_{32} &=
\begin{pmatrix} 
0 \\ 
1 \\
\mu_c+y_k^{[l-1]}\\
\mu_c^2 + 2\sigma^{*2}_{l-1}(\mathbf{x}) + \left(y_k^{[l-1]}\right)^2 + \mu_c y_k^{[l-1]}\\
\mu_c^3 + \left(y_k^{[l-1]}\right)^3 + y_k^{[l-1]} \mu_c \left( \mu_c +y_k^{[l-1]} \right) + \sigma^{*2}_{l-1}(\mathbf{x}) \left( 5\mu_c + 3y_k^{[l-1]} \right)
\end{pmatrix}, 
\end{align*}

\begin{align*}
\Lambda_{41} &= 
\begin{pmatrix} 
1 \\ 
\mu^*\\
\mu^{*2}+\sigma^{*2}_{l-1}(\mathbf{x})\\
\mu^* \left( \mu^{*2}+3\sigma^{*2}_{l-1}(\mathbf{x}) \right) \\
\mu^{*4} + 6\mu_c^2\sigma^{*2}_{l-1}(\mathbf{x}) + 3\sigma^{*4}_{l-1}(\mathbf{x})
\end{pmatrix}, \mu^*=\mu^*_{l-1}(\mathbf{x}),
\\
\Lambda_{42} &=
\begin{pmatrix} 
0 \\ 
1 \\
\mu^*+y_i^{[l-1]}\\
\mu^{*2} + 2\sigma^{*2}_{l-1}(\mathbf{x}) + \left(y_i^{[l-1]}\right)^2 + \mu^*y_i^{[l-1]}\\
\mu^{*3} + \left(y_i^{[l-1]}\right)^3 + y_i^{[l-1]} \mu^* \left( \mu^* + y_i^{[l-1]} \right) + \sigma^{*2}_{l-1}(\mathbf{x}) \left( 5\mu^* + 3y_i^{[l-1]} \right)
\end{pmatrix}, \\
\Lambda_{43} &=
\begin{pmatrix} 
0 \\ 
1 \\
\mu^*+y_k^{[l-1]}\\
\mu^{*2} + 2\sigma^{*2}_{l-1}(\mathbf{x}) + \left(y_k^{[l-1]}\right)^2 + \mu^*y_k^{[l-1]}\\
\mu^{*3} + \left(y_k^{[l-1]}\right)^3 + y_k^{[l-1]} \mu^* \left( \mu^* + y_k^{[l-1]} \right) + \sigma^{*2}_{l-1}(\mathbf{x}) \left( 5\mu^* + 3y_k^{[l-1]} \right)
\end{pmatrix}, \\
\Lambda_{51} &= 
\begin{pmatrix} 
1 \\ 
-\mu_d\\
\mu_d^2+\sigma^{*2}_{l-1}(\mathbf{x})\\
-\mu_d \left( \mu_d^2+3\sigma^{*2}_{l-1}(\mathbf{x}) \right) \\
\mu_d^4 + 6\mu_d^2\sigma^{*2}_{l-1}(\mathbf{x}) + 3\sigma^{*4}_{l-1}(\mathbf{x})
\end{pmatrix}, \mu_d=\mu^*_{l-1}(\mathbf{x}) + 2\sqrt{5}\frac{\sigma^{*2}_{l-1}(\mathbf{x})}{\theta_{ly}},
\\
\Lambda_{52} &=
\begin{pmatrix} 
0 \\ 
1 \\
-\mu_d-y_i^{[l-1]}\\
\mu_d^2 + 2\sigma^{*2}_{l-1}(\mathbf{x}) + \left(y_i^{[l-1]}\right)^2 + \mu_d y_i^{[l-1]}\\
-\mu_d^3 - \left(y_i^{[l-1]}\right)^3 - y_i^{[l-1]} \mu_d \left( \mu_d + y_i^{[l-1]} \right) - \sigma^{*2}_{l-1}(\mathbf{x}) \left( 5\mu_d + 3y_i^{[l-1]} \right)
\end{pmatrix}, 
\end{align*}

\begin{align*}
E_{1} &= \frac{1}{3\theta_{ly}^2}
\begin{pmatrix} 
3\theta_{ly}^2 - 3\sqrt{5}\theta_{ly}y_i^{[l-1]} + 5 \left(y_i^{[l-1]}\right)^2 \\ 
3\sqrt{5}\theta_{ly} - 10y_i^{[l-1]}\\
5
\end{pmatrix},
E_{2} = \frac{1}{3\theta_{ly}^2}
\begin{pmatrix} 
3\theta_{ly}^2 + 3\sqrt{5}\theta_{ly}y_i^{[l-1]} + 5 \left(y_i^{[l-1]}\right)^2 \\ 
3\sqrt{5}\theta_{ly} + 10y_i^{[l-1]}\\
5
\end{pmatrix},\\
E_{3} &= \frac{1}{9\theta_{ly}^4}
\begin{pmatrix} 
E_{31} &
E_{32} &
E_{33} &
E_{34} &
E_{35} 
\end{pmatrix} ^ \top,\\
E_{31} &= 9\theta_{ly}^4 + 25\left( y_i^{[l-1]} \right)^2 \left( y_k^{[l-1]} \right)^2 -3\sqrt{5}\theta_{ly} \left( 3\theta_{ly}^2 + 5 y_i^{[l-1]} y_k^{[l-1]} \right)\left( y_i^{[l-1]} + y_k^{[l-1]} \right)\\
&+15 \theta_{ly}^2 \left( \left( y_i^{[l-1]} \right)^2 + \left( y_k^{[l-1]} \right)^2 + 3y_i^{[l-1]}y_k^{[l-1]}  \right) , \\
E_{32} &= 18\sqrt{5} \theta_{ly}^3 + 15\sqrt{5} \theta_{ly} \left( \left( y_i^{[l-1]} \right)^2 + \left( y_k^{[l-1]} \right)^2 \right) -75 \theta_{ly}^2 \left( y_i^{[l-1]} + y_k^{[l-1]} \right) \\
&-50 y_i^{[l-1]} y_k^{[l-1]}  \left( y_i^{[l-1]} + y_k^{[l-1]} \right) + 60\sqrt{5} \theta_{ly} y_i^{[l-1]} y_k^{[l-1]} ,\\
E_{33} &= 5 \left\{ 5\left( y_i^{[l-1]} \right)^2 + 5\left( y_k^{[l-1]} \right)^2 + 15\theta_{ly}^2 - 9\sqrt{5} \theta_{ly} \left( y_i^{[l-1]} + y_k^{[l-1]} \right) + 20 \left( y_i^{[l-1]} y_k^{[l-1]} \right) \right\} \\
E_{34} &= 10 \left( 3\sqrt{5} \theta_{ly} - 5 y_i^{[l-1]}  - 5 y_k^{[l-1]} \right) , E_{35} = 25,\\
E_{4} &= \frac{1}{9\theta_{ly}^4}
\begin{pmatrix} 
E_{41} &
E_{42} &
E_{43} &
E_{44} &
E_{45} 
\end{pmatrix} ^ \top,\\
E_{41} &= 9\theta_{ly}^4 + 25\left( y_i^{[l-1]} \right)^2 \left( y_k^{[l-1]} \right)^2  +3\sqrt{5}\theta_{ly} \left( 3\theta_{ly}^2 - 5 y_i^{[l-1]} y_k^{[l-1]} \right)\left( y_i^{[l-1]} - y_k^{[l-1]} \right)\\
&+15 \theta_{ly}^2 \left( \left( y_i^{[l-1]} \right)^2 + \left( y_k^{[l-1]} \right)^2 - 3 y_i^{[l-1]} y_k^{[l-1]}  \right) , \\
E_{42} &= 5 \left\{ 3\sqrt{5} \theta_{ly} \left( \left( y_k^{[l-1]} \right)^2 - \left( y_i^{[l-1]} \right)^2 \right) + 3 \theta_{ly}^2 \left( y_i^{[l-1]} + y_k^{[l-1]} \right) -10 y_i^{[l-1]} y_k^{[l-1]} \left( y_i^{[l-1]} + y_k^{[l-1]} \right) \right\},\\
E_{43} &= 5 \left\{ 5\left( y_i^{[l-1]} \right)^2 + 5\left( y_k^{[l-1]} \right)^2 -3 \theta_{ly}^2 - 3\sqrt{5} \theta_{ly} \left( y_k^{[l-1]} - y_i^{[l-1]} \right) + 20 \left( y_i^{[l-1]} y_k^{[l-1]} \right) \right\} \\
E_{44} &= -50 \left(  y_i^{[l-1]} + y_k^{[l-1]} \right) , E_{45} = 25,\\
E_{5} &= \frac{1}{9\theta_{ly}^4}
\begin{pmatrix} 
E_{51} &
E_{52} &
E_{53} &
E_{54} &
E_{55} 
\end{pmatrix} ^ \top,\\
E_{51} &= 9\theta_{ly}^4 + 25\left( y_i^{[l-1]} \right)^2 \left( y_k^{[l-1]} \right)^2 +3\sqrt{5}\theta_{ly} \left( 3\theta_{ly}^2 + 5 y_i^{[l-1]} y_k^{[l-1]} \right)\left( y_i^{[l-1]} + y_k^{[l-1]} \right)\\
&+15 \theta_{ly}^2 \left( \left( y_i^{[l-1]} \right)^2 + \left( y_k^{[l-1]} \right)^2 + 3y_i^{[l-1]} y_k^{[l-1]}  \right) , \\
E_{52} &= 18\sqrt{5} \theta_{ly}^3 + 15\sqrt{5} \theta_{ly} \left( \left( y_i^{[l-1]} \right)^2 + \left( y_k^{[l-1]} \right)^2 \right) +75 \theta_{ly}^2 \left( y_i^{[l-1]} + y_k^{[l-1]} \right) \\
&+50 y_i^{[l-1]} y_k^{[l-1]}  \left( y_i^{[l-1]} + y_k^{[l-1]} \right) + 60\sqrt{5} \theta_{ly} y_i^{[l-1]} y_k^{[l-1]} ,
\end{align*}

\begin{align*}
E_{53} &= 5 \left\{ 5\left( y_i^{[l-1]} \right)^2 + 5\left( y_k^{[l-1]} \right)^2 + 15\theta_{ly}^2 + 9\sqrt{5} \theta_{ly} \left( y_i^{[l-1]} + y_k^{[l-1]} \right) + 20 \left( y_i^{[l-1]} y_k^{[l-1]} \right) \right\} \\
E_{54} &= 10 \left( 3\sqrt{5} \theta_{ly} + 5 y_i^{[l-1]} + 5 y_k^{[l-1]} \right) , E_{55} = 25,
\end{align*}
for $ y_i^{[l-1]} \leq y_k^{[l-1]}$. If $ y_i^{[l-1]} > y_k^{[l-1]}$, interchange $ y_i^{[l-1]}$ and $ y_k^{[l-1]}$.

\section{Proof of Proposition \ref{prop:interpolation}}\label{app:proof3.2}
For $l=1$, since $\mu^*_1(\mathbf{x})=\mu_1(\mathbf{x})$ and $\sigma^{*2}_1(\mathbf{x})=\sigma^{2}_1(\mathbf{x})$ are the posterior mean and variance of a conventional GP, it can be shown that $\mu^*_1(\mathbf{x}^{[1]}_i)=y^{[1]}_i$ and $\sigma^{*2}_1(\mathbf{x}^{[1]}_i)=0$ \citep{suppsantner2018design}. For $l=k-1$, suppose that $\mu^*_{k-1}(\mathbf{x}^{[k-1]}_i)=y^{[k-1]}_i$ and $\sigma^{*2}_{k-1}(\mathbf{x}^{[k-1]}_i)=0$, implying that $f_{k-1}(\mathbf{x}^{[k-1]}_i)|\mathbf{y}_1,\ldots,\mathbf{y}_{k-1}$ remains constant at the value  $y^{[k-1]}_i$. Then, 
\begin{align*}
\mu^*_k(\mathbf{x}_i^{[k]}) =&\mathbb{E}[f_k(x_i^{[k]})|\mathbf{y}_1,\ldots,\mathbf{y}_k]
= \mathbb{E} [ \mathbb{E} [f_k(\mathbf{x}_i^{[k]}) | f_{k-1}(\mathbf{x}_i^{[k]}), \mathbf{y}_1,\ldots,\mathbf{y}_k ] ]\\
=&\mathbb{E}[\mu_k(\mathbf{x}^{[k]}_i,f_{k-1}(\mathbf{x}_i^{[k]}))|\mathbf{y}_1,\ldots,\mathbf{y}_k]\\
=&\mu_k(\mathbf{x}^{[k]}_i,y^{[k-1]}_i)=y^{[k]}_i,
\end{align*}
and
\begin{align*}
\sigma^{*2}_k(\mathbf{x}_i^{[k]})&=\mathbb{V}[f_k(\mathbf{x}_i^{[k]})|\mathbf{y}_1,\ldots,\mathbf{y}_k] \\
&= \mathbb{V} [ \mathbb{E} [f_k(\mathbf{x}_i^{[k]}) | f_{k-1}(\mathbf{x}_i^{[k]}), \mathbf{y}_1,\ldots,\mathbf{y}_k ] ]+\mathbb{E} [ \mathbb{V} [f_k(\mathbf{x}_i^{[k]}) | f_{k-1}(\mathbf{x}_i^{[k]}), \mathbf{y}_1,\ldots,\mathbf{y}_k ] ] \\
&= \mathbb{V}[\mu_k(\mathbf{x}^{[k]}_i,f_{k-1}(\mathbf{x}_i^{[k]}))|\mathbf{y}_1,\ldots,\mathbf{y}_k] + \mathbb{E}[\sigma^2_k(\mathbf{x}^{[k]}_i,f_{k-1}(\mathbf{x}_i^{[k]}))|\mathbf{y}_1,\ldots,\mathbf{y}_k]\\
&=0+\sigma^2_k(\mathbf{x}^{[k]}_i,y^{[k-1]}_i)=0,
\end{align*}
where $\mu^2_k(\mathbf{x},f_{k-1}(\mathbf{x}))$ and $\sigma^2_k(\mathbf{x},f_{k-1}(\mathbf{x}))$ are defined in \eqref{eq:gppostmean2} and \eqref{eq:gppostvar2}. 
In both of the derivations, the second to last equation holds because of  $\mathbf{x}^{[k]}_i=\mathbf{x}^{[k-1]}_i$ and $f_{k-1}(\mathbf{x}^{[k-1]}_i)|\mathbf{y}_1,\ldots,\mathbf{y}_{k-1}:=y^{[k]}_i$. The last equation holds because of the interpolation property of conventional GPs. By induction, this finishes the proof.

\section{Synthetic functions in Section \ref{sec:emulationperformance}}\label{app:functions}

\subsection{Two-level Perdikaris function} 
\begin{align*}
\begin{cases}
 &f_1(x) = \sin(8 \pi x) \\
 &f_2(x) = (x-\sqrt{2})f_1(x)^2
\end{cases} 
\text{for } x \in [0,1].
\end{align*}

\subsection{Two-level Park function} 
\begin{align*}
\begin{cases}
 &f_1(\mathbf{x}) = f_2(\mathbf{x})+\frac{\sin(x_1)}{10} f_2(\mathbf{x}) -2x_1+x_2^2+x_3^2+0.5 \\
 &f_2(\mathbf{x}) = \frac{x_1}{2} \left[ \sqrt{1+(x_2+x_3^2)\frac{x_4}{x_1^2}} -1 \right] +(x_1+3x_4) \exp{(1+\sin(x_3))}
\end{cases}  
\text{for } \mathbf{x} \in [0,1]^4.
\end{align*}

\subsection{Three-level Branin function}
\begin{align*}
\begin{cases}
 &f_1(\mathbf{x}) = f_2(1.2(\mathbf{x}+2)) -3x_2 +1 \\
 &f_2(\mathbf{x}) = 10\sqrt{f_3(\mathbf{x})} +2(x_1-0.5) -3(3x_2-1) -1 \\
 &f_3(\mathbf{x}) = \left( \frac{-1.275 x_1^2}{\pi^2} +\frac{5x_1}{\pi} +x_2-6 \right)^2 + \left( 10-\frac{5}{4\pi} \right) \cos(x_1) +10
\end{cases}  
\text{for } \mathbf{x} \in [-5,10] \times [0,15].
\end{align*}  

\subsection{Two-level borehole function}
\begin{align*}
\begin{cases}
 &f_1(\mathbf{x}) = \frac{2\pi T_u(H_u-H_l)}{\ln (r/r_w)(1+\frac{2LT_u}{\ln (r/r_w)r^2_wK_w}+\frac{T_u}{T_l})} \\
 &f_2(\mathbf{x}) = \frac{5 T_u(H_u-H_l)}{\ln (r/r_w)(1.5+\frac{2LT_u}{\ln (r/r_w)r^2_wK_w}+\frac{T_u}{T_l})},
\end{cases}  
\end{align*}
where $r_w\in[0.05, 0.15],r\in[100, 50000],T_u\in[63070, 115600],H_u\in[990, 1110],T_l\in[63.1, 116],H_l\in[700, 820],L\in[1120, 1680]$ and $K_w\in[9855, 12045]$. 

\subsection{Two-level Currin function}
\begin{align*}
\begin{cases}
 &f_1(\mathbf{x})=\frac{1}{4}\left[f_2(x_1+0.05,x_2+0.05)+f_2(x_1+0.05,\max(0,x_2-0.05))\right]\\
 &\quad\quad\quad+\frac{1}{4}\left[f_2(x_1-0.05,x_2+0.05)+f_2(x_1-0.05,\max(0,x_2-0.05))\right]\\
 &f_2(\mathbf{x}) = \left[1-\exp\left(-\frac{1}{2x_2}\right)\right]\frac{2300x_1^3 + 1900x_1^2 + 2092x_1 + 60}{100x_1^3 + 500x_1^2 + 4x_1 + 20}
\end{cases}  
\text{for } \mathbf{x} \in [0,1]^2.
\end{align*}

\subsection{Three-level Franke function}

\begin{align*}
\begin{cases}
 &f_1(\mathbf{x}) = 0.75\exp\left(-\frac{(9x_1-2)^2}{4}-\frac{(9x_2-2)^2}{4}\right)+0.75\exp\left(-\frac{(9x_1+1)^2}{49}-\frac{9x_2+1}{10}\right) \\
 &\quad\quad+0.5\exp\left(-\frac{(9x_1-7)^2}{4}-\frac{(9x_2-3)^2}{4}\right)-0.2\exp\left(-(9x_1-4)^2-(9x_2-7)^2\right)\\
 &f_2(\mathbf{x}) = \exp(-1.4f_1(\mathbf{x}))\cos(3.5\pi f_1(\mathbf{x})) \\
 &f_3(\mathbf{x}) =\sin(2\pi(f_2(\mathbf{x})-1))
\end{cases}  
\text{for } \mathbf{x} \in [0,1]^2.
\end{align*}

\section{Variance decomposition}\label{supp:variancedecomposition}
In this section, we detail the calculation of $V_l(\mathbf{x})$ in (15) for the settings of $L=2$ and $L=3$ with the squared exponential kernel function (7).

Denote $\mathbf{Y}_l=\{\mathbf{y}_1,\ldots,\mathbf{y}_l\}$. In the setting of $L=2$, it follows that $\sigma^{*2}_2(\mathbf{x})=V_1(\mathbf{x})+V_2(\mathbf{x})$ with
\begin{align}\label{eq:V1}
V_1(\mathbf{x})=&\mathbb{V} \left[ \mathbb{E} [f_2(\mathbf{x}) | f_1(\mathbf{x}), \mathbf{Y}_2 ] \right]\nonumber\\
=& \mathbb{E} \left[ \left\{ \mathbb{E} \left[f_2(\mathbf{x}) | f_1(\mathbf{x}), \mathbf{y}_1, \mathbf{y}_2 \right] \right\}^2 \right] - \mu^{*}_2(\mathbf{x})^2 \nonumber \\
=& - (\mu^*_2(\mathbf{x})-\alpha_2)^2 + \mathbb{E}\left[ \left\{ \mathbf{k}_2 (\mathbf{x},f_1(\mathbf{x}))^T  \mathbf{K}^{-1}_2  (\mathbf{y}_2 - \alpha_2 \mathbf{1}_{n_2}) \right\}^2 |\mathbf{Y}_2\right] \nonumber \\
=& - (\mu^*_2(\mathbf{x})-\alpha_2)^2 \nonumber \\
&+ \sum_{i,k=1}^{n_2} \frac{r^{{[2]}}_ir^{{[2]}}_k}{\sqrt{1+4\frac{\sigma^{*2}_1(\mathbf{x})}{\theta_{2y}}}}\exp{ \left(-\sum^d_{j=1}\frac{(x_j-x^{[2]}_{ij})^2+(x_j-x^{[2]}_{kj})^2}{\theta_{2j}}-a^{[2]}_{ik}(\mathbf{x})\right)} ,
\end{align}  
\begin{align}\label{eq:V2}
V_2(\mathbf{x})=&\mathbb{E} \left[\mathbb{V} [f_2(\mathbf{x}) | f_1(\mathbf{x}), \mathbf{Y}_2 ] \right]\nonumber\\
=& \mathbb{E} \left[ \tau_2^2 \left\{ 1 - \mathbf{k}_2 (\mathbf{x},f_1(\mathbf{x}))^T  \mathbf{K}^{-1}_2  \mathbf{k}_2(\mathbf{x},f_1(\mathbf{x})) \right\} |\mathbf{Y}_2 \right] \nonumber \\
=&\tau^2_2\left(1 - \frac{1}{\sqrt{1+4\frac{\sigma^{*2}_1(\mathbf{x})}{\theta_{2y}}}}\sum_{i,k=1}^{n_2} (\mathbf{K}^{-1}_2)_{ik}  \exp\left( -\sum^d_{j=1}\frac{(x_j-x^{[2]}_{ij})^2+(x_j-x^{[2]}_{kj})^2}{\theta_{2j}}-a^{[2]}_{ik}(\mathbf{x})\right) \right),
\end{align}     
where $$r_i^{{[2]}} = (\mathbf{K}^{-1}_2 (\mathbf{y}_2-\alpha_2\mathbf{1}_{n_2}))_i\quad\text{,}\quad
a^{[l]}_{ik}(\mathbf{x})=\frac{(\frac{y^{[l-1]}_i+y^{[l-1]}_k}{2}-\mu^*_{l-1}(\mathbf{x}))^2}{\frac{\theta_{ly}}{2}+2\sigma^{*2}_{l-1}(\mathbf{x})} +\frac{(y^{[l-1]}_i-y^{[l-1]}_k)^2}{2\theta_{ly}}.$$

In the setting of $L=3$, it follows that 
\begin{align*}
\sigma^{*2}_3(\mathbf{x})=&\mathbb{V}[f_3(\mathbf{x})|\mathbf{Y}_3]=\mathbb{V}[\mathbb{E}[f_3(\mathbf{x})|f_2(\mathbf{x}),\mathbf{Y}_3]]+\mathbb{E}[\mathbb{V}[f_3(\mathbf{x})|f_2(\mathbf{x}),\mathbf{Y}_3]]\\
=&\mathbb{V}\mathbb{E}[\mathbb{E}[f_3(\mathbf{x})|f_2(\mathbf{x}),f_1(\mathbf{x}),\mathbf{Y}_3]]+\mathbb{E}\mathbb{V}[\mathbb{E}[f_3(\mathbf{x})|f_2(\mathbf{x}),f_1(\mathbf{x}),\mathbf{Y}_3]]+\mathbb{E}\mathbb{E}[\mathbb{V}[f_3(\mathbf{x})|f_2(\mathbf{x}),f_1(\mathbf{x}),\mathbf{Y}_3]]\\
:=&V_1(\mathbf{x})+V_2(\mathbf{x})+V_3(\mathbf{x}).
\end{align*}
Each contribution $V_l(\mathbf{x})$ is calculated as follows. For $V_1(\mathbf{x})$, it follows that 
\begin{align*}
V_1(\mathbf{x})=&\mathbb{V}\mathbb{E}\mathbb{E}[f_3(\mathbf{x}) | f_2(\mathbf{x}), f_1(\mathbf{x}), \mathbf{Y}_3 ] \\
=&\mathbb{V} \mathbb{E} [ \alpha_3 +\mathbf{k}_3 (\mathbf{x},f_2(\mathbf{x}))^T \mathbf{K}^{-1}_3 (\mathbf{y}_3 - \alpha_3 \mathbf{1}_{n_3}) | f_1(\mathbf{x}), \mathbf{Y}_3 ] \\
=&\mathbb{V} \left[ \sum^{n_3}_{i=1} r^{[3]}_i \prod_{j=1}^d \exp\left( -\frac{(x_{j}-x^{[3]}_{ij})^2}{\theta_{3j}} \right)  \frac{1}{\sqrt{1+2\frac{
\sigma^{2}_{2}(\mathbf{x},f_1(\mathbf{x})) }{\theta_{3y}}}}  \exp{\left( -\frac{(y_i^{[2]}-\mu_{2}(\mathbf{x},f_1(\mathbf{x})))^2}{\theta_{3y}+2\sigma^{2}_{2}(\mathbf{x},f_1(\mathbf{x}))} \right)} \mid\mathbf{Y}_3 \right] 
\end{align*}
where $r_i^{{[3]}} = (\mathbf{K}^{-1}_3 (\mathbf{y}_3-\alpha_3\mathbf{1}_{n_3}))_i$, and $\mu_{2}(\mathbf{x},f_1(\mathbf{x}))$ and $\sigma^{2}_{2}(\mathbf{x},f_1(\mathbf{x}))$ are defined in (10) and (11), respectively. Since $f_1(\mathbf{x})|\mathbf{y}_1\sim\mathcal{N}(\mu_1(\mathbf{x}),\sigma_1^2(\mathbf{x}))$ with $\mu_1(\mathbf{x})$  and $\sigma_1^2(\mathbf{x})$ given in (8) and (9), respectively, $V_1(\mathbf{x})$ can be easily approximated using Monte-Carlo methods. For example, let $J(\mathbf{x},f_1(\mathbf{x}))=\sum^{n_3}_{i=1} r^{[3]}_i \prod_{j=1}^d \exp\left( -\frac{(x_{j}-x^{[3]}_{ij})^2}{\theta_{3j}} \right)  \frac{1}{\sqrt{1+2\frac{
\sigma^{2}_{2}(\mathbf{x},f_1(\mathbf{x})) }{\theta_{3y}}}}  \exp{\left( -\frac{(y_i^{[2]}-\mu_{2}(\mathbf{x},f_1(\mathbf{x})))^2}{\theta_{3y}+2\sigma^{2}_{2}(\mathbf{x},f_1(\mathbf{x}))} \right)}$ and $\{\tilde{y}_i\}^N_{i=1}$ are the $N$ random samples from the normal distribution $\mathcal{N}(\mu_1(\mathbf{x}),\sigma_1^2(\mathbf{x}))$, then 
$$
V_1(\mathbf{x})\approx\frac{1}{N}\sum^N_{i=1}\left(J(\mathbf{x},\tilde{y}_i)-\frac{1}{N}\sum^N_{k=1}J(\mathbf{x},\tilde{y}_k)\right)^2.
$$

For $V_2(\mathbf{x})$, it follows that 
\begin{align*}
V_2(\mathbf{x})=&\mathbb{E}  \mathbb{V} \mathbb{E}[f_3(\mathbf{x}) | f_2(\mathbf{x}), f_1(\mathbf{x}), \mathbf{Y}_3 ] \\
=&\mathbb{E} \mathbb{E}\left[ \left\{ \mathbb{E} \left[f_3(\mathbf{x}) | f_2(\mathbf{x}),f_1(\mathbf{x}), \mathbf{Y}_3 \right] \right\}^2 \right] - \mu^{*}_3(\mathbf{x})^2\\
=& - (\mu^*_3(\mathbf{x})-\alpha_3)^2 \nonumber \\
& + \sum_{i,k=1}^{n_3} r^{{[3]}}_ir^{{[3]}}_k\prod^d_{j=1}\exp{ \left(-\frac{(x_j-x^{[3]}_{ij})^2+(x_j-x^{[3]}_{kj})^2}{\theta_{3j}}\right)} \mathbb{E}\left[ b^{[3]}_{ik}(\mathbf{x},f_1(\mathbf{x}))|\mathbf{Y}_3\right], 
\end{align*}
where $$b^{[3]}_{ik}(\mathbf{x},f_1(\mathbf{x}))=\frac{1}{\sqrt{1+4\frac{\sigma^{2}_2(\mathbf{x}, f_1(\mathbf{x}))}{\theta_{3y}}}}  \exp{\left( -\frac{(\frac{y_i^{[2]}+y_k^{[2]}}{2}-\mu_2(\mathbf{x}, f_1(\mathbf{x})))^2}{\frac{\theta_{3y}}{2}+2\sigma^{2}_{2}(\mathbf{x}, f_1(\mathbf{x}))} -\frac{(y_i^{[2]}-y_k^{[2]})^2}{2\theta_{3y}} \right)}.$$
Similarly, $\mathbb{E}\left[ b^{[3]}_{ik}(\mathbf{x},f_1(\mathbf{x}))|\mathbf{Y}_3\right] $ can be approximated by 
$$
\mathbb{E}\left[ b^{[3]}_{ik}(\mathbf{x},f_1(\mathbf{x}))|\mathbf{Y}_3\right] \approx\frac{1}{N}\sum^N_{i=1}b^{[3]}_{ik}(\mathbf{x},\tilde{y}_i).
$$

Lastly, for $V_3(\mathbf{x})$, it follows that
\begin{align*}
V_3(\mathbf{x})=&\mathbb{E} \mathbb{E} \mathbb{V}[f_3(\mathbf{x}) | f_2(\mathbf{x}), f_1(\mathbf{x}), \mathbf{Y}_3] =\mathbb{E} \mathbb{V}[f_3(\mathbf{x}) | f_2(\mathbf{x}), \mathbf{Y}_3]\\
=& \mathbb{E} \left[ \tau_3^2 \left\{ 1 - \mathbf{k}_3 (\mathbf{x},f_2(\mathbf{x}))^T  \mathbf{K}^{-1}_3  \mathbf{k}_3(\mathbf{x},f_2(\mathbf{x})) \right\} | \mathbf{Y}_3 \right]  \nonumber \\
=&\tau^2_3\left(1 - \frac{1}{\sqrt{1+4\frac{\sigma^{*2}_2(\mathbf{x})}{\theta_{3y}}}}\sum_{i,k=1}^{n_3} (\mathbf{K}^{-1}_3)_{ik}  \exp\left( -\sum^d_{j=1}\frac{(x_j-x^{[3]}_{ij})^2+(x_j-x^{[3]}_{kj})^2}{\theta_{3j}}-a^{[3]}_{ik}(\mathbf{x})\right) \right).
\end{align*}

\section{Definition of RMSE and CRPS}\label{supp:RMSE}
The root-mean-square error (RMSE) and continuous rank probability score (CRPS) are defined as follows:
\begin{align*}
    \text{RMSE} &=  \left(\sum_{i=1}^{n_{\rm{test}}} \frac{\left( f_L(\mathbf{x}^{\rm{test}}_i) - \mu^*_L(\mathbf{x}^{\rm{test}}_i) \right)^2}{n_{\rm{test}}}\right)^{1/2}  ,\\
    \text{CRPS} &= \frac{1}{n_{\rm{test}}}\sum^{n_{\rm{test}}}_{i=1}\sigma_L^*(\mathbf{x}^{\rm{test}}_i) \left[ \frac{1}{\sqrt{\pi}} - 2\phi \left(z(\mathbf{x}^{\rm{test}}_i) \right) - z(\mathbf{x}^{\rm{test}}_i) \left( 2\Phi(z(\mathbf{x}^{\rm{test}}_i)) -1 \right)  \right],
\end{align*}
with $z(\mathbf{x}^{\rm{test}}_i)=(f_L(\mathbf{x}^{\rm{test}}_i) - \mu^*_L(\mathbf{x}^{\rm{test}}_i))/\sigma_L^*(\mathbf{x}^{\rm{test}}_i)$, where $\mathbf{x}^{\rm{test}}_i$ is the $i$-th test input location, $\phi$ and $\Phi$ denote the probability density function and cumulative distribution function of a standard normal distribution, respectively.

\section{Supporting tables and figures in Sections 5 and 6}\label{supp:figures}

\begin{table}[h]
    \centering
    \begin{tabular}{c|C{4.5em}|C{4.5em}|C{4.5em}|C{4.5em}|C{4.5em}|C{4.5em}}
        & Perdikaris & Branin& Park& Borehole& Currin& Franke \\
        \hline
         $d$ &1& 2& 4& 8&2&2 \\
         $n_1$ & 13&20& 40&60&20&20 \\
         $n_2$ & 8&15& 20&30&10&15 \\
         $n_3$ &&10&&&&10 \\
    \end{tabular}
    \caption{Dimension and sample sizes for each of the synthetic examples.}
    \label{tab:simulationstudy}
\end{table}

\begin{figure}[h]
\begin{center}
\includegraphics[width=0.9\textwidth]{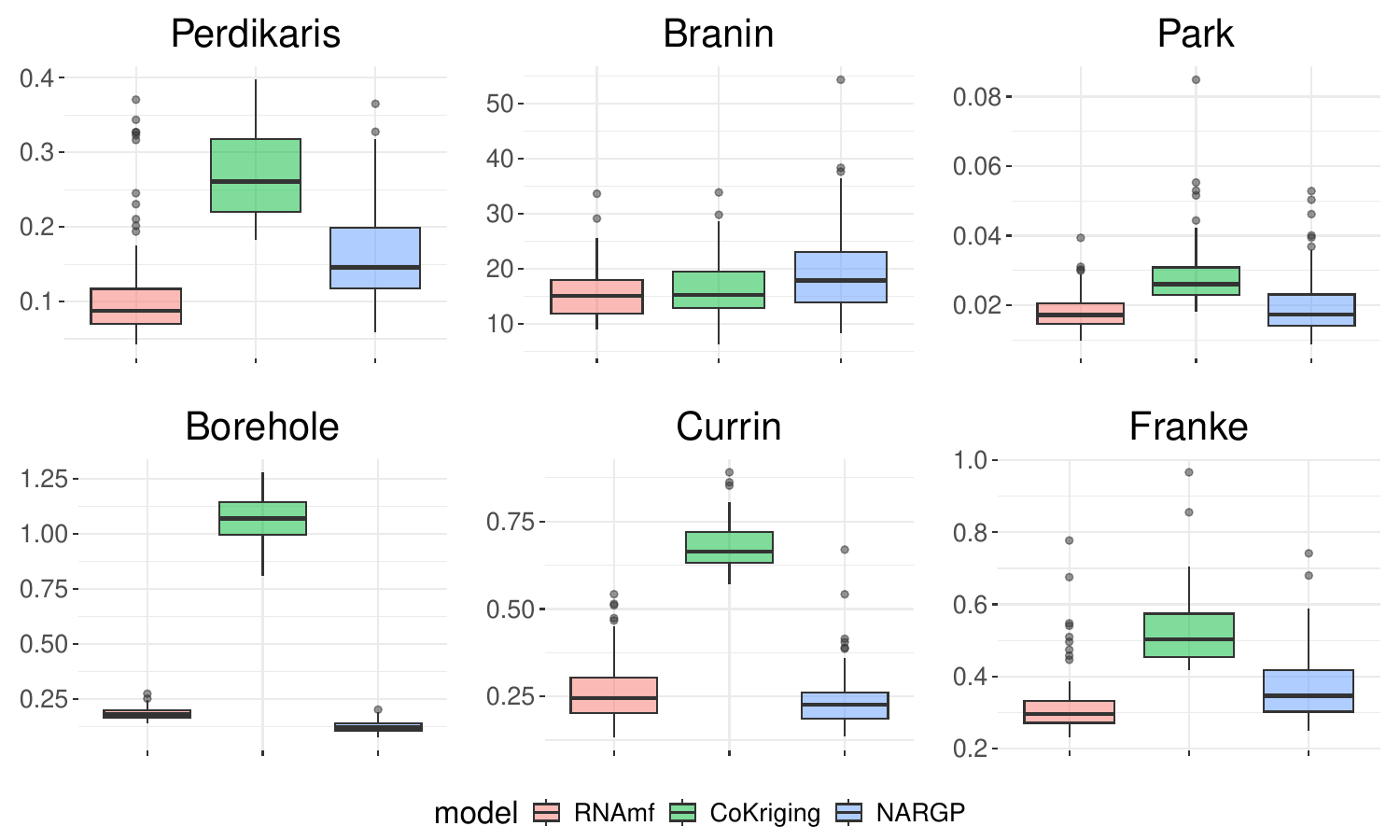} 
\end{center}
\caption{CRPSs of six synthetic examples in Section 5.1 across 100 repetitions.}
\label{fig:num_CRPS}
\end{figure}

\begin{figure}[h]
\begin{center}
\includegraphics[width=\textwidth]{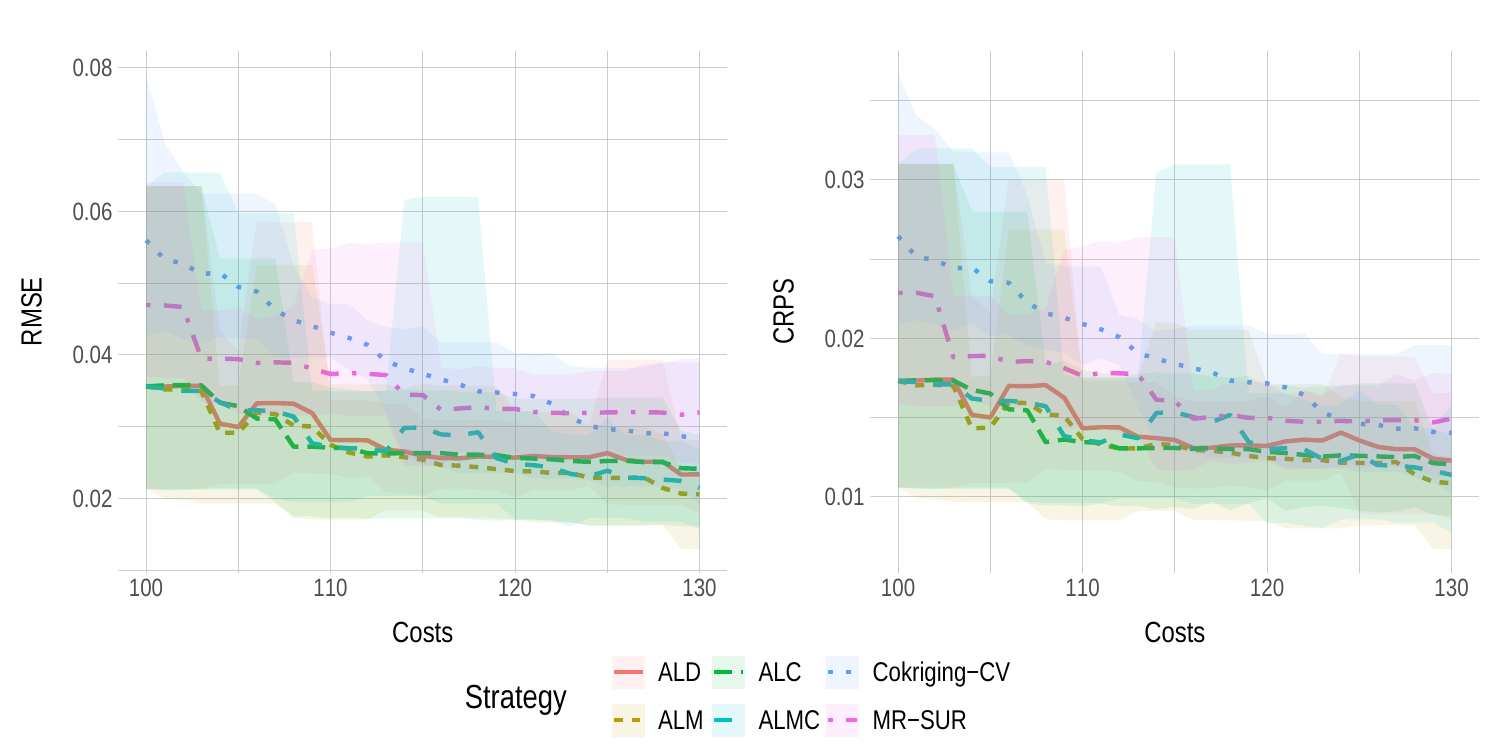} 
\end{center}
\caption{RMSE and CRPS for the Park function with respect to the simulation cost. Solid lines represent the average over 10 repetitions and shaded regions represent the ranges.}
\label{fig:AC_park}
\end{figure}

\begin{figure}[h]
\begin{center}
\includegraphics[width=0.9\textwidth]{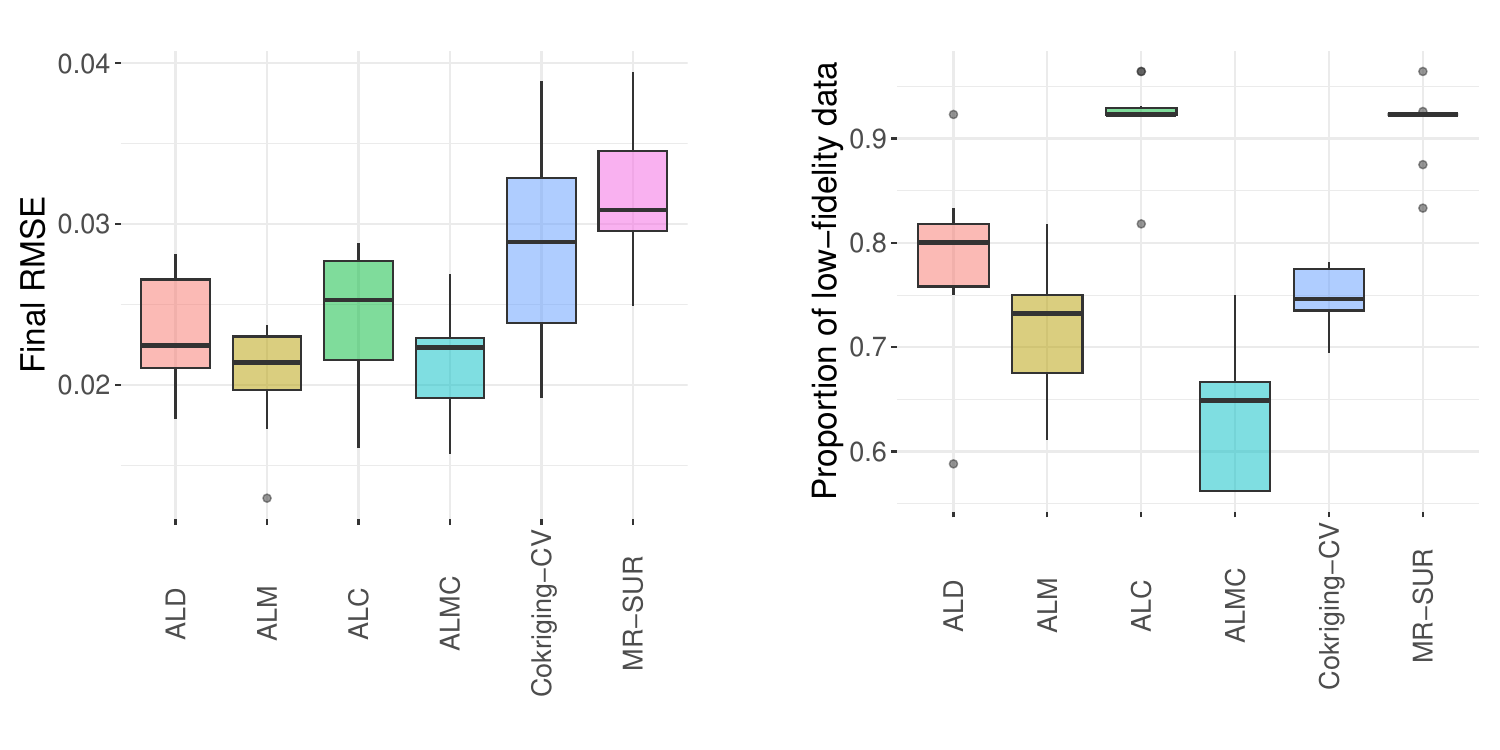} 
\end{center}
\caption{Final RMSE (left) and proportion of AL acquisitions choosing low-fidelity data (right) for the Park function. Boxplots indicate spread over 10 repetitions.}
\label{fig:AC_park_proportion}
\end{figure}

\begin{figure}[h]
\begin{center}
\includegraphics[width=0.9\textwidth]{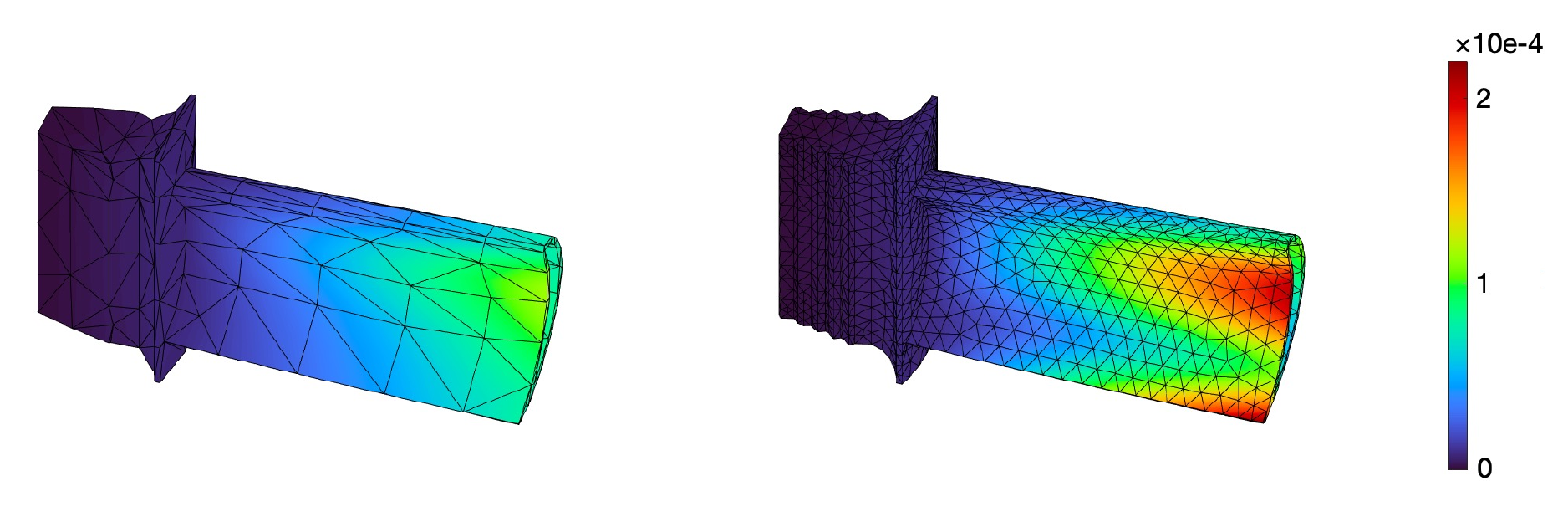} 
\end{center}
\caption{
Illustration of low-fidelity (left) and high-fidelity (right) finite element simulations at the input setting $\mathbf{x}=(0.5,0.45)$ in the turbine blade application.}
\label{fig:turbine}
\end{figure}

\end{document}